\documentclass[useAMS,usenatbib]{mn2e}

\usepackage{graphicx}
\usepackage{color}
\usepackage{adjustbox}

\newbox\grsign \setbox\grsign=\hbox{$>$} \newdimen\grdimen \grdimen=\ht\grsign
\newbox\simlessbox \newbox\simgreatbox
\setbox\simgreatbox=\hbox{\raise.5ex\hbox{$>$}\llap
     {\lower.5ex\hbox{$\sim$}}}\ht1=\grdimen\dp1=0pt
\setbox\simlessbox=\hbox{\raise.5ex\hbox{$<$}\llap
     {\lower.5ex\hbox{$\sim$}}}\ht2=\grdimen\dp2=0pt
\def\simgreater{\mathrel{\copy\simgreatbox}}
\def\simless{\mathrel{\copy\simlessbox}}
\newbox\simppropto
\setbox\simppropto=\hbox{\raise.5ex\hbox{$\sim$}\llap
     {\lower.5ex\hbox{$\propto$}}}\ht2=\grdimen\dp2=0pt

\title[N-Rich Stars in Inner Galaxy]{Chemical tagging with APOGEE:
Discovery of a large population of N-rich stars in the inner Galaxy}
\author[Schiavon et al.]
{Ricardo P. Schiavon$^{1}$\thanks{E-mail: R.P.Schiavon@ljmu.ac.uk},
Olga Zamora$^{2,3}$, 
Ricardo Carrera$^{2,3}$,
\newauthor
Sara Lucatello$^{4}$,
A.C. Robin$^{5}$,
Melissa Ness$^{6}$,
Sarah L. Martell$^{7}$,
\newauthor
Verne V. Smith$^{8,9}$,
D.~A. Garc\'\i a Hern\'andez$^{2,3}$,
Arturo Manchado$^{2,3,10}$,
\newauthor
Ralph Sch\"onrich$^{11}$,
Nate Bastian$^{1}$,
Cristina Chiappini$^{12}$,
Matthew Shetrone$^{13}$,
\newauthor
J. Ted Mackereth$^{1}$,
Rob A. Williams$^{1}$,
Szabolcs M\'esz\'aros$^{14}$,
\newauthor
Carlos Allende Prieto$^{2,3}$,
Friedrich Anders$^{10}$,
Dmitry Bizyaev$^{15,16}$,
\newauthor
Timothy C. Beers$^{17}$,
S. Drew Chojnowski$^{18}$,
Katia Cunha$^{8,9}$,
\newauthor
Courtney Epstein$^{20}$,
Peter M. Frinchaboy$^{21}$,
Ana E. Garc\'\i a P\'erez$^{2}$,
\newauthor
Fred R. Hearty$^{22}$,
Jon A. Holtzman$^{23}$,
Jennifer A. Johnson$^{20}$,
\newauthor
Karen Kinemuchi$^{15}$,
Steven R. Majewski$^{18}$,
Demitri Muna$^{20}$,
\newauthor
David L. Nidever$^{24,25,26}$,
Duy Cuong Nguyen$^{27}$,
Robert W. O'Connell$^{18}$,
\newauthor
Daniel Oravetz$^{15}$,
Kaike Pan$^{15}$,
Marc Pinsonneault$^{20}$,
\newauthor
Donald P. Schneider$^{22}$,
Matthias Schultheis$^{28}$,
Audrey Simmons$^{15}$,
\newauthor
Michael F. Skrutskie$^{18}$,
Jennifer Sobeck$^{18}$,
John C. Wilson$^{18}$
\& Gail Zasowski$^{29}$
\\
\\
$^{1}$Astrophysics Research Institute, Liverpool John Moores University,
146 Brownlow Hill, Liverpool, L3 5RF, United Kingdom \\
$^{2}$Instituto de Astrof\'\i sica de Canarias, E-38205 La Laguna, Tenerife,
Spain\\
$^{3}$Departamento de Astrof\'\i sica, Universidad de La Laguna (ULL),
E-38206 La Laguna, Tenerife, Spain\\
$^{4}$INAF-Osservatorio Astronomico di Padova, Vicolo dell’Osservatorio 5,
	I-35122 Padova, Italy \\
$^{5}$Institut Utinam, CNRS UMR6213, Universit\'e de Franche-Comt\'e, OSU
THETA Franche-Comt\'e-Bourgogne, \\
~~Observatoire de Besan\c con, BP 1615, 25010 Besan\c con Cedex, France\\
$^{6}$Max-Planck-Institut fur Astronomie, K\"onigstuhl 17, D-69117
Heidelberg, Germany\\
$^{7}$School of Physics, University of New South Wales, Sydney, NSW 2052,
Australia\\
$^{8}$National Optical Astronomy Observatories, Tucson, AZ 85719, USA \\
$^{9}$Steward Observatory, University of Arizona, Tucson, AZ 85721, USA\\
$^{10}$Consejo Superior de Investigaciones Cientificas (CSIC)\\
$^{11}$Rudolf Peierls Centre for Theoretical Physics, 1 Keble Road, Oxford,
OX1 3NP, UK\\
$^{12}$Leibniz-Institut f\"ur Astrophysik Potsdam (AIP), An der Sternwarte
16, 14482 Potsdam, Germany \\
$^{13}$University of Texas at Austin, McDonald Observatory, Fort Davis, TX
79734, USA\\
$^{14}$ELTE Gothard Astrophysical Observatory, H-9704 Szombathely, Szent
Imre Herceg st. 112, Hungary\\
$^{15}$Apache Point Observatory, P.O. Box 59, Sunspot, NM 88349-0059, USA \\
$^{16}$Sternberg Astronomical Institute, Moscow State University, Moscow \\
$^{17}$Dept. of Physics and JINA Center for the Origin of the Elements, 
University of Notre Dame Notre Dame, IN  46556  USA\\
$^{18}$Dept. of Astronomy, University of Virginia, Charlottesville,
	VA 22904-4325, USA \\
$^{19}$Observat\'orio Nacional, S\~ao Crist\'ov\~ao, Rio de Janeiro,
Brazil \\
$^{20}$Department of Astronomy, The Ohio State University, Columbus,
	OH 43210, USA \\
$^{21}$Texas Christian University, Fort Worth, TX 76129, USA \\
$^{22}$Department of Astronomy and Astrophysics, 
Institute for Gravitation and the Cosmos, The Pennsylvania State University, \\
University Park, PA 16802 \\
$^{23}$New Mexico State University, Las Cruces, NM 88003, USA \\
$^{24}$Department of Astronomy, University of Michigan, Ann Arbor, MI,
48104, USA\\
$^{25}$Large Synoptic Survey Telescope, 950 North Cherry Ave, Tucson, AZ
85719 \\
$^{26}$Steward Observatory, 933 North Cherry Ave, Tucson, AZ 85719 \\
$^{27}$Dunlap Institute for Astronomy and Astrophysics, University
	of Toronto, Toronto, Ontario, Canada  \\
$^{28}$Laboratoire Lagrange (UMR7293), Universite de Nice Sophia Antipolis,
CNRS, Observatoire de la C ´ ote d’Azur, \\
~~BP 4229, F-06304 Nice Cedex 4,
France\\
$^{29}$Department of Physics and Astronomy, Johns Hopkins University, 
Baltimore, MD 21218, USA
}
\begin{document}

\date{Fourth draft, 11 September, 2015}

\pagerange{\pageref{firstpage}--\pageref{lastpage}} \pubyear{2015}

\maketitle

\label{firstpage}

\clearpage

\begin{abstract}

Formation of globular clusters (GCs), the Galactic bulge, or
galaxy bulges in general, are important unsolved problems
in Galactic astronomy.  Homogeneous infrared observations
of large samples of stars belonging to GCs and the
Galactic bulge field are one of the best ways to study these problems.
We report the discovery by APOGEE of a population of
field stars in the inner Galaxy with abundances of N, C, and Al
that are typically found in GC stars.  The newly
discovered stars have high [N/Fe], which is correlated with [Al/Fe]
and anti-correlated with [C/Fe].  They are homogeneously distributed
across, and kinematically indistinguishable from, other field stars
in the same volume.  Their metallicity distribution 
is seemingly unimodal, peaking at [Fe/H]$\sim$--1, thus being in
disagreement with that of the Galactic GC system.
Our results can be understood in terms of different scenarios.
N-rich stars could be former members of dissolved GCs,
in which case the mass in destroyed GCs exceeds that
of the surviving GC system by a factor of $\sim$8.  In that scenario,
the total mass contained in so-called ``first-generation'' stars cannot be larger
than that in ``second-generation'' stars by more than a factor of
$\sim$9 and was certainly smaller.  Conversely, our results
may imply the absence of a mandatory genetic
link between ``second generation'' stars and
GCs.  Last, but not least, N-rich stars could
be the oldest stars in the Galaxy, the by-products of chemical
enrichment by the first stellar generations formed in the heart of the
Galaxy.



\end{abstract}


\section{Introduction} \label{intro}

While central to our understanding of the formation of the Galaxy,
the birthplaces of the stars that make up its main components are
not well known.  In the case of the Galactic halo, this classical
problem \citep{els,sz78} has been framed in modern times within the
context of galaxy formation theory \citep{wr78,bl84} in a $\Lambda$-CDM
universe \citep{sp03}.  Recent evidence that the Galactic halo is
split into an inner and an outer component, with distinct chemical
compositions \citep{ca07,car10,an15,fe15} goes along with theoretical predictions
for the origin of those components \citep[e.g.,][]{mc12,ti14}, at least
in a qualitative sense \citep[although see][for an alternative
view]{sc14}.  Most importantly, both data and models have reached
a degree of sophistication that allows one to begin addressing
detailed questions about the nature of the original star forming
units that gestated the stars seen in the halo today---in particular
their characteristic masses \citep[e.g.,][]{fi15,de15}.  

Regarding the bulge, the situation is considerably less clear, which
is due partly to difficult observational access to the inner Galaxy,
and partly to the short dynamical timescales, which caused signatures
of the early stellar systems to be erased from phase space long
ago.  Moreover, the physical overlap of all components of the
Galaxy (halo, thin and thick disks, bar, and bulge) within its inner
few kpc makes a definition of the pertinence of a given star or
stellar group to any of those components quite difficult, making
the very definition of the bulge itself somewhat contentious.
The literature on the Galactic bulge is sufficiently vast to render
any attempt at a summary here quite vain.  However, the picture
emerging from even a brief examination of the state of the art is
that of a current lack of a unique definition of the nature of the
bulge, both in terms of the distribution of its components in phase
space and, to a lesser extent, in terms of its stellar population
content.  We therefore use the term ``bulge'' somewhat loosely,
without necessarily implying the existence of a classical spheroidal
structure detached from the inner halo, but simply referring to the
aggregate of all stellar mass cohabiting the central few kpc of the
Galaxy.

Studies of stellar bulge spatial distribution, kinematics, metallicity
distribution function, metallicity gradients, and abundance patterns
paint a complex picture.  Several groups have confirmed the
presence of a complex metallicity distribution of bulge stellar
populations \citep[e.g.,][]{zo08,hi11,jo11,ne13a,de13,ra14}, with
the presence of multiple components, each with characteristic
structure and kinematics. At high metallicity ([Fe/H]~$\simgreater$~--0.5),
the bulge appears to be dominated by a boxy/peanut-shaped structure,
associated with a bar \citep{bs91}, which in projection has been
found to assume an X shape in 2MASS maps \citep{mz10,ne12}.  On the
other hand, stars in the low-metallicity end take on a more spheroidal
distribution and are thought to be associated with either the thick
disk or halo \citep[e.g.,][]{ba10,ne13a,ne13b,ra14} or perhaps even
a classical spheroidal bulge \citep[e.g.,][]{ba10,hi11}, although
the existence of the latter has been called into question by model
fits to stellar counts in the 2MASS and SDSS catalogs \citep{ro14}.
Evidence from kinematics pointing to the existence of an anti-correlation
between metallicity and velocity dispersion \citep[e.g.,][]{jo11,jo14}
jibes well with the above picture.  More detailed studies, based
on high quality radial velocities for samples of many thousand stars
show that, while the metal-rich bulge population rotates cylindrically,
the kinematics of metal-poor stars is consistent with a slowly
rotating spheroid, possibly due to a combination of thick disk and
halo \citep{ne13b,ne15}.  Finally, detailed chemical abundance
studies showed, since the early work by \cite{mr94}, that metal-poor
stars tend to be $\alpha$-enhanced, as in the thick disk and halo,
whereas their metal-rich counterparts have [$\alpha$/Fe] close to
solar \citep[see also][]{hi11,ne13a,jo14,ry16}.  When studied
individually, $\alpha$ elements were found not to follow all exactly
the same trend with metallicity \citep{fu07} and, perhaps most
importantly, to be slightly more enhanced in metal-rich bulge stars
than in their thin and thick disk counterparts \citep{jo14}.  

One possible approach to gain insights into the nature of star
forming units that gave origin to the stars in the Galaxy is through
chemical tagging \citep{fb02,ti15a,ti15b}.  The method consists of
using very detailed and accurate chemical compositions to identify
stars sharing a common origin, with the hope of tracing them back
to their original star forming units.  In principle, chemical tagging
can be extremely powerful, provided that each and every star forming
unit was characterised by a {\it unique} detailed abundance pattern.
If this assumption is correct, one could distinguish every single
star forming unit from all others by determining a large enough
number of elemental abundances.  That, of course, is observationally
very costly.  A weaker, less expensive, form of the method consists
of associating a given abundance pattern not necessarily to a unique
star forming unit, but to an entire class---say, stellar clusters
above a given mass.  In order to work, this {\it weak chemical
tagging} requires knowledge of a smaller number of elemental
abundances than does standard strong chemical tagging.


An early application of weak chemical tagging was pursued by
\cite{mg10} and \cite{ma11} \citep[see also][]{ca10,li15}.  These authors
discovered halo field stars with very high nitrogen and relatively
low carbon abundances, which is an abundance pattern characteristic
of some particular globular cluster (GC) populations.  On the theory
that these stars were originally formed in GCs, Martell and
collaborators concluded that they resulted from the dissolution of
GCs a claim that is in line with detections of tidal tails
around GCs such as Palomar 5 \citep{od03} and NGC 5466 \citep{be06a}.
By accounting for the expected fraction of GC stars with normal N
abundances, \cite{ma11} estimated that at least 17\% of the stellar
mass in the Galactic halo resulted from the dissolution of GCs
and/or their parent systems.  Based on the same results, but adopting
a different set of assumptions, \cite{gr12} estimated that {\it
most} of the halo has in fact originated from those systems.
Following yet another approach based on a model of the chemical and
dynamical evolution of the Galactic GC system, \cite{sc11} estimate
that up to 10--20\% of the Galactic halo mass was contributed by tidal
evaporation of Galactic GCs.

One major difference in these estimates is that they assign different
theoretically motivated ratios between the numbers of ``enriched''
and ``normal'' stars.  Our lack of a firm handle on the origin of
the multiple-population phenomenon in Galactic GCs is therefore an
important limiting factor.  Moreover, uncertainties about the shape
of the initial mass function of the Galactic GC system also play a
role.  The presence of large spreads of elemental abundances, and
anti-correlations thereof, in GC stars has been long known
\citep[e.g.,][]{nz77,nc79,dc80} and consistently confirmed by more
recent observations \citep[e.g.,][]{ca09,me15}.\footnote{
For reviews on star-to-star abundance variations in GCs, see
\cite{gr04} and \cite{gr12}.} In the past decade, abundance spreads
were ascribed to the clusters' intrinsic chemical evolution, either
due to some form of feedback-regulated star formation history
\citep[e.g.,][]{de07,de08,re08,cs11} or to other processes
\citep[e.g.,][]{ba13,ho14}.  However, none of the existing models
put forth so far can account for the existing chemical composition
data in detail \citep{ba15}.  Yet, the different models make vastly
different predictions.  In particular, models that propose chemical
evolution through feedback-regulated star formation postulate that
GCs were 10-100 times more massive in the past \citep[for
discussion and references, see, e.g.,][]{gr12,bl15,cz15}. Moreover,
according to these models, the vast majority of this mass must have
been in the form of {\it first generation}\footnote{The
widespread use of the term ``generation'' to refer to each of the
multiple populations in GCs is associated with the sometimes tacit
acceptance of a specific set of scenarios for their origin.  Despite
the fact that existing models do not account for the extant
data in detail, which calls into question the physical reality of
these scenarios, we choose to adopt the same nomenclature for
consistency with current jargon.} stars (henceforth, FG stars) which so
far remain chemically indistinguishable from field stars of the same
[Fe/H].  These conditions are required so that, for any
reasonable initial mass function, early stellar generations can
produce the necessary amounts of light elements observed in {\it
second generation stars} (SG stars)---those with
enhanced He, N, Na, and Al abundances.
This issue is referred to as the {\it mass budget problem} \citep{re08}.



In this paper, we report the serendipitous discovery of a population
of bulge field stars with abundance patterns that are similar to those
found in stars from globular clusters.  Characterised by high [N/Fe],
which is correlated with [Al/Fe] and anti-correlated with [C/Fe],
these stars are homogeneously distributed across the Galactic bulge
and, to first order, are spatially and kinematically indistinguishable
from the rest of the bulge field population.  We characterise this
new population and discuss the implications of this finding for our
understanding of both bulge and GC formation.

In Section~\ref{data} the data employed in this paper are described.
The results are presented in Section~\ref{results} and discussed
in Section~\ref{discussion}.  Our conclusions are summarised in
Section~\ref{summary}.

\section{Data and sample} \label{data}

The results reported in this paper are based on elemental abundances
for a large sample of Galactic stars from Data Release 12
\citep[DR12,][]{al15} of the Apache Point Observatory Galactic
Evolution Experiment \citep[APOGEE,][]{ma15}.  One of four Sloan
Digital Sky Survey-III \citep[SDSS-III,][]{ei11} experiments, APOGEE
used a new spectrograph on the Sloan 2.5~m telescope \citep{gu06}
at APO to obtain high quality H-band spectra (R=22,500, S/N $\sim$
100 per half-resolution element) for over 136,000 stars distributed
across all Galactic components, from which precision radial velocities,
stellar parameters, and abundances for up to 15 elements have been
obtained.  Further in-depth information on the APOGEE survey, data,
and the data reduction pipeline can be found in \cite{ma15},
\cite{ho15}, and \cite{ni15}, respectively.  The APOGEE Stellar
Parameters and Chemical Abundances Pipeline (ASPCAP) is described
in detail in \cite{gp15a}.  Heliocentric distances, $d_\odot$, 
were based on a Bayesian analysis of the stellar parameters, adopting
as priors a history of star formation and initial mass function,
and the PARSEC theoretical isochrones \citep{br12}.  For more details
on the method, see \cite{bi14}.  Possible systematic effects in our
distances were assessed by employing the linear distance estimator
from \cite{sc12}, with the conclusion that distances to the giants
may be too long by about 25\% in general, with indications for
possibly a larger systematic effect at $\log g < 2$, which does not
affect our results.

In this paper, we concern ourselves
with a subset of the APOGEE sample, namely stars located in the
Galactic bulge.  Moreover, because we are interested in searching
for stars with chemical signatures typical of GC members, we focus
on the abundances of Fe, C, N, and Al.  With the above constraints
in mind, the sample analysed in this study is defined by the following
set of criteria:

\begin{enumerate}
\item $|b| < 16^\circ$
\item $-20^\circ < l < 20^\circ$
\item 5 kpc $< d_\odot <$ 11 kpc
\item 3500 K $<$ $T_{\rm eff}$ $<$ 4500 K
\item $\log g < 3.6$
\item S/N $>$ 70 pixel$^{-1}$
\end{enumerate}

The final sample so selected amounts to a total of 5,148 stars.
Because we are interested in field stars only, suspected or
known members of GCs located within the spatial region defined above
were identified and excluded from the sample.  A star was considered
to be a GC member if it is located within its tidal radius, if its
radial velocity differs from that of the GC (when available) by no
more than 20 $km s^{-1}$, and if its metallicity differs from that
of the cluster by no more than 0.3 dex.  In this way 8 stars were
identified as GC members, within the range of distances above,
leaving us with a grand total of 5,140 field stars.  The surface
gravity criterion is meant to avoid contamination of the sample by
nearby dwarfs---which are in any case extremely rare in APOGEE bulge
pointings, given the shallower magnitude limit adopted in these
fields---see \cite{za13} for details.  The $T_{\rm eff}$ criterion
is aimed at maximizing the overall quality of the abundances
considered.  At $T_{\rm eff} < 3500$~K, APOGEE does not presently
provide elemental abundances, because the spectral library upon
which ASPCAP is based does not extend to such low temperatures.  At
the other end of the $T_{\rm eff}$ range, stars hotter than 4500~K
are not considered because the abundances of C and N are uncertain
in that $T_{\rm eff}$ regime \citep[see discussion in][]{me15}.
The uncertainty arises because ASPCAP determines these abundances
from the strengths of CN and CO lines, which become too weak for
$T_{\rm eff} > 4500$~K at relatively low metallicities ([Fe/H]
$\simless$--1).  This sample is supplemented with data for stars
belonging to various Galactic GCs targeted by APOGEE \citep[for
details, see][]{za13,me15}, that meet the selection criteria on
stellar parameters and S/N listed above.  The latter data set is
used to define the locus occupied by GC stars in chemical diagnostic
plots.  

The DR12 APOGEE abundances employed in this work are based
on $\chi^2$-minimisation of the observed spectra against a large
spectral library calculated on the basis of state of the art model
photospheres \citep{me12} and a customised line list \citep{sh15}.
Specifically, synthetic spectra were calculated using the
ASS$\varepsilon$T code \citep{ko09}, using LTE, plane parallel,
model photospheres calculated with the ATLAS9 code \citep{ku93}.
Giant stars with $T_{\rm eff} \simless 4000~K$ and low surface
gravity present extended atmospheres, which can invalidate the
plane-parallel approximation.  Sphericity effects cause a dilution
of radiative flux that leads to lower temperatures in the upper
layers of the photosphere \citep{pl92}, potentially affecting the
strengths of molecular lines.  To check for any important
systematics coming from adoption of plane-parallel photospheres,
DR12 abundances were compared with those obtained from a run of
ASPCAP adopting a new spectral library \citep{za15}, calculated
using the Turbospectrum spectrum synthesis code \citep{ap98,pl12}
and the MARCS model atmospheres \citep{gu08}, which adopt spherical
symmetry for all models with $\log g \leq 3$.  These comparisons
showed that the elemental abundances relevant to this work are not
affected by adoption of those more sophisticated analysis methods,
so we proceed by adopting DR12 numbers for the remainder of this
study.

A cautionary note is in order before proceeding with the analysis.
The elemental abundances from APOGEE are subject to zero-point
differences relative to optical studies.  Calibrations between the
APOGEE and literature abundance scales were performed by \cite{me13}
on DR10 data \citep{ah14}, whereas, for DR12, a similar procedure
was followed, as described by \cite{ho15}.  In this study, unless
otherwise noted, we opt to work with the uncorrected data, to take
full advantage of the homogeneity and internal consistency of the
ASPCAP-derived elemental abundances, which is crucial when comparing
samples from different systems, such as the Galactic bulge and GCs.
As discussed by \cite{ho15}, the zero-point corrections are small
and do not affect our conclusions in any important way.

\section{Results}  \label{results}

In this section we describe the central finding reported in this
paper, namely, the discovery of stars in the field of the inner
Galaxy that possess chemical compositions that are suggestive of a
globular cluster origin.  In Sections 3.1 and 3.2 the behaviour of
the sample in chemical composition diagnostic plots is characterised,
and the identification of the newly discovered stellar population
is described.  Section 3.3 is aimed at reassuring the reader of the
reality of the high nitrogen abundances resulting from ASPCAP.
Sections 3.4 and 3.5 discuss the possible contamination of our
sample by other stars that could potentially present the same
abundance patterns, respectively intermediate-mass AGB stars and
the secondary remnants of mass-transfer binaries, concluding that
such contaminations are minimal and not likely to affect our results
importantly.  The reader solely interested in the discussion
of the main results may skip the latter three sub-sections.


\subsection{Distribution of field stars in the [Fe/H]-[N/Fe] plane}
\label{distrib}

\begin{figure}
 \centering
 \includegraphics[width=100mm]{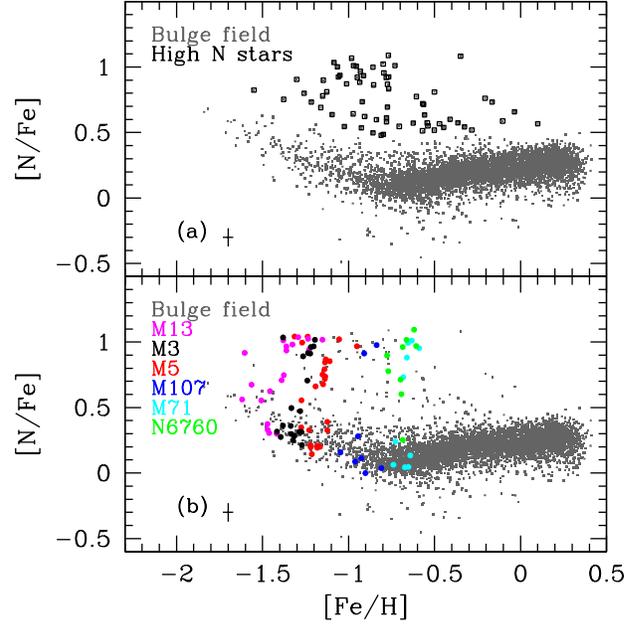}
 \caption{Distribution of the 5,140 sample stars in the [N/Fe] vs
 [Fe/H] plane.  {\it (a)} Shown with small gray dots are 
 stars selected as described in Section~\ref{data}.  Squares
 indicate N-rich stars, defined as stars deviating by more than
 4$\sigma$ from a $6^{th}$ order polynomial fit to the relation
 between [N/Fe] and [Fe/H].  {\it (b)} The same field sample
 is compared to APOGEE data for Galactic globular clusters within
 a smaller range of metallicities.  Globular cluster stars span the
 range of [N/Fe] covered by field stars of same metallicity.
 First-generation GC stars are a good match to the
 [N/Fe] vs [Fe/H] relation of field stars, as expected \citep[see,
 e.g.,][]{gr12}, with SG stars spanning larger values
 of [N/Fe] for fixed [Fe/H], thus occupying the same locus as the
 N-rich stars reported in this paper.  The sharp edge in the data
 distribution at [N/Fe] $\sim$ 1 is an artefact of the
 [N/Fe] upper limit in the ASPCAP spectral library.
 }
 \label{nfe_bulge}
\end{figure}

We start by examining the distribution of our sample stars in the
[Fe/H] vs [N/Fe] plane, which is presented in Figure~\ref{nfe_bulge}a.
Three main features are worth noticing in this plot.  First, 
is clear that the relation between [N/Fe] and [Fe/H] in 
of the sample is non-monotonic.  The bulk of the stars with
[Fe/H]$\simgreater$--0.7 follow a clear correlation between [Fe/H]
and [N/Fe].  Second, for lower metallicities there is a reversal
in that relation, such that [N/Fe] actually decreases with increasing
[Fe/H].  Third, a large number of stars, highlighted by
adoption of larger symbols are scattered above the main swath of
data points at all metallicities, and there is a smaller number of
outliers towards low [N/Fe] values.  A total of 67 high-[N/Fe]
outliers, highlighted by larger symbols, are identified by fitting
a $6^{th}$ order polynomial to the [N/Fe] vs. [Fe/H] relation and
selecting stars that deviate from the fit by more than 4$\sigma$.
For reasons that are explained in Section~\ref{nrich}, we remove
stars with [C/Fe]$>$+0.15, leaving a total sample of 58 stars, which
we henceforth refer to as {\it N-rich stars}.  They are listed
in Table~\ref{list}.



Post main-sequence evolution complicates the interpretation
of carbon and nitrogen abundances in giant stars \citep[see,
e.g.,][]{la12}.  The surface abundances of these elements are
affected by the combined effects of the first dredge-up and extra
mixing \citep[for a thorough review, see][]{kl14}.  The first
dredge-up is a well understood physical process that involves the
deepening of the convection zone as the star evolves up the giant
branch, causing material processed through the CN(O) cycle to be
brought to the stellar surface, changing the atmospheric abundances
of some elements.  Extra mixing, on the other hand, is a nonconvective
process that brings about additional changes to atmospheric abundances
in red giants \citep[e.g.,][]{gr00}.  The efficiency of extra mixing
is a function of stellar metallicity \citep{ma08}, and the physical
process responsible for it has not been established yet.  Some of
the ideas proposed involve stellar rotation, thermohaline mixing,
magnetic fields, meridional circulation combined with turbulent
diffusion, or perhaps some combination of some of these processes
\citep[e.g.,][]{rv81,cl10,an12,kl14}.  An in-depth analysis of
mixing is beyond the scope of this paper.  While deep mixing hampers
interpretation of these data in terms of the history of nitrogen
and carbon enrichment of the Galaxy, it has no impact on our
results, as discussed in Section~\ref{nrich}.


\begin{figure}
 \centering
 \includegraphics[width=100mm]{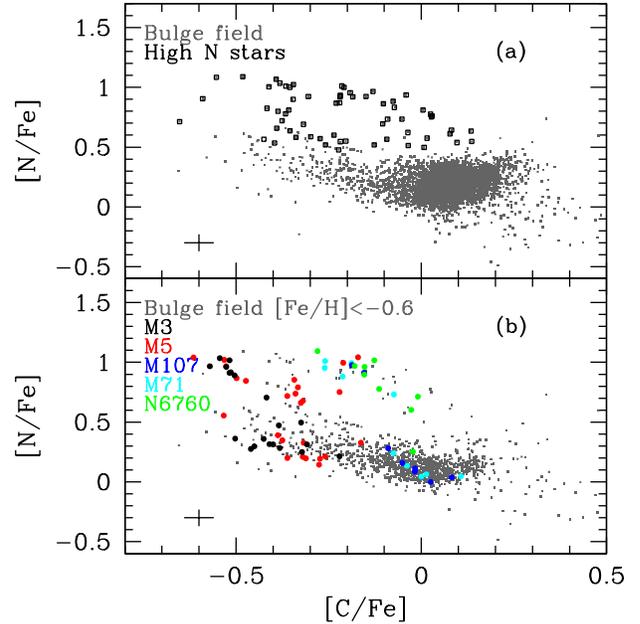}
 \caption{Stars from Figure~\ref{nfe_bulge} in the [C/Fe] vs [N/Fe]
 plane.  {\it (a)} This is the figure where
 a population of stars with high nitrogen abundance, anti-correlated 
 with carbon (N-rich stars), was first identified.  There potentially
 are two populations of N-rich stars, with [N/Fe] up to $\sim$1, and
 another with lower [N/Fe] and a smaller range of [C/Fe] (roughly 
 between --0.5 and --0.2).  Stars marked as squares
 are the same as those in Figure~\ref{nfe_bulge}, except for those
 with [C/Fe] $>$ +0.15, because those seem to depart from the
 anti-correlation between C and N.  Moreover, stars with such high
 [C/Fe] are not usually found in Galactic globular clusters.  {\it
 (b)} Bulge field stars are compared with members of Galactic GCs.
 The GC members are distributed across discrete ``branches''
 within which the two abundances are anti-correlated.  Each branch belongs
 to a distinct GC stellar population, and the anti-correlation
 within each branch is due to stellar evolution.  In each GC,
 branches with the lowest [N/Fe] abundances correspond to FG
 populations, whose abundance patterns are indistinguishable
 from the bulk of field stars of same metallicity.  GC branches with higher
 [N/Fe] correspond to SG stars, which occupy the same locus 
 in the C-N plan as N-rich stars, showing an identical anti-correlation between 
 [C/Fe] and [N/Fe].  Bulge field
 stars with [Fe/H]$<$--0.6 and M13 members are excluded, to keep
 the comparison to stars within the same metallicity range.}
 \label{nc_bulge}
\end{figure}



\begin{figure}
 \centering
 \includegraphics[width=100mm]{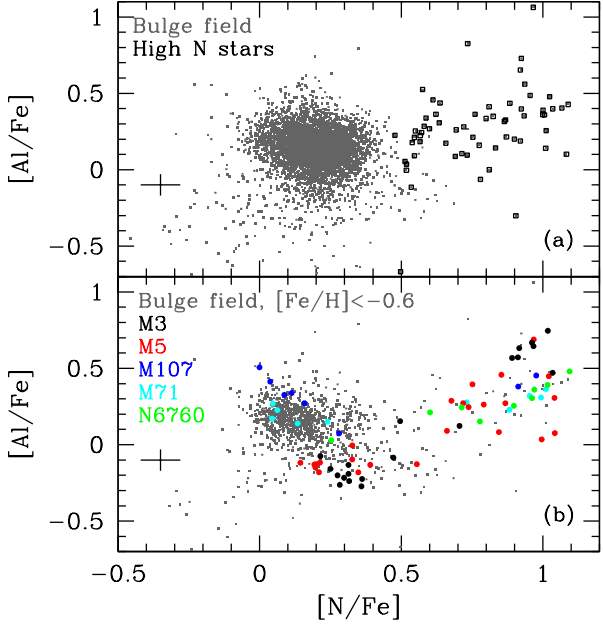}
 \caption{Sample stars in another chemical abundance diagnostic plot,
adopting same symbols as in Figures~\ref{nc_bulge}a,b.  {\it (a)} A
correlation between [N/Fe] and [Al/Fe] is evident in the N-rich sample
(squares), whereas an
anti-correlation appears to be present for stars with normal nitrogen
abundances ([N/Fe]$\simless+0.5$).  {\it (b)} The SG stars in GCs,
with [N/Fe] $\simgreater$ +0.5, occupies the same locus as the
N-rich population, following a similar correlation between the two
abundance ratios.  On the other hand, FG stars in GCs follow
the same trend as the
the lower [N/Fe] field stars.  This plot corroborates the notion that
the N-rich stars inhabit the same region of chemical composition space as
SG stars from GCs.  The GC sample has few counterparts
to the field stars with intermediate-nitrogen ($0.5 \simless$ [N/Fe]
$\simless0.7$ in panel {\it (a)}.  It is unclear whether this is a real
effect or whether it is due to limitations in the APOGEE GC sample, which is 
relatively small.}
\label{aln_bulge}
\end{figure}

\subsection{N-rich stars}  \label{nrich}

Now we turn our attention to the high [N/Fe] outliers in
Figure~\ref{nfe_bulge}a.  A diagnostic plot that can shed light on
the nature of those stars is shown in Figure~\ref{nc_bulge}a where
N-rich stars are again highlighted by large symbols.  One can
see that the N-rich stars are distributed along at least two discrete
bands where [N/Fe] is anti-correlated with [C/Fe].  The most obvious
of these branches contains the stars with highest [N/Fe] in our
sample, and runs roughly between \{[C/Fe],[N/Fe]\} = \{--0.5,+1.1\}
and \{0.0,+0.8\}.  There also seems to be a second branch of stars
with more intermediate values of [N/Fe], running approximately between
\{[C/Fe],[N/Fe]\} = \{--0.4,+0.8\} and \{--0.1,+0.4\}.  This
intermediate-N branch is less obvious and partly merged with the
main body of the stars with [C/Fe] $\simless$ --0.1.  Nevertheless,
close inspection of Figure~\ref{nc_bulge} reveals a fairly clear
intermediate-N sequence, which is well separated from the main body
of N-normal stars for [C/Fe] $<$ -0.2.

These individual branches of anti-correlated carbon and nitrogen
abundances strongly resemble those long known to exist in Galactic
GCs \citep[e.g.,][]{no81,he82}.  As mentioned above, the mixing of
the products of high-temperature shell hydrogen burning into the
atmospheres of the giant branch stars leads to a large increase in
the relative abundance of nitrogen along the giant branch, at the
expense of a comparable, (or smaller, depending on the cluster or
data set), decrease in the relative abundance of carbon.  Over time,
RGB stars evolve along these branches, upward and to the left through
the [C/Fe]-[N/Fe] plane.  The similarity with GC stars is further
suggested by Figure~\ref{nc_bulge}b, where APOGEE data for stars
belonging to a few Galactic GCs, spanning the same range of
metallicities as the N-rich stars, are overlaid on the field sample.
This plot indicates that the GC member stars follow similar C-N
anti-correlation sequences that run parallel to those observed in
the field.  In this Figure, GCs show clear evidence for the presence
of more than one C-N anticorrelation sequence.  Each of these
sequences corresponds to one of the multiple populations commonly
found in Galactic globular clusters \citep[see][for a discussion
of the APOGEE data on these clusters]{me15}.

Since the stars in both field and GC samples are red giants that
have undergone first dredge-up, their nitrogen abundances have
increased from their initial main-sequence values at the expense
of $^{12}$C.  Expected variations are a function of stellar mass
and, to a lesser degree, metallicity, but [N/Fe] typically increases
by $\sim$ 0.3 dex \citep[e.g.,][]{cl10}, whereas [C/Fe] decreases
by $\sim-0.15$.  The stars belonging to the ``normal'' sequence
reach [N/Fe] as high as $\sim$+0.5 and [C/Fe] around $\sim-0.5$,
which may be extreme values for first dredge-up in low-mass red
giants, but can be accounted for by efficient deep mixing during
the first-ascent giant branch.  Taking the example of a star with
[Fe/H]=$-1.0$ and the observed trend in Figure~\ref{nfe_bulge}, the
expected initial $^{14}$N abundance would be [N/Fe]~$\sim+0.1$, or
A(N)$\sim$6.8 for a solar nitrogen value of A$_{\odot}$=7.86.  At
the same [Fe/H], normal stars in our sample have an average value
of $\sim-0.15$ for [C/Fe], which corresponds to A(C)$\sim$7.25.  If
most of this initial carbon ($^{12}$C) is converted to $^{14}$N via
the CN-cycle, then the red giant nitrogen abundance could approach
$7.4 - 7.5$, or [N/Fe] around +0.5 or +0.6.  Such values of $^{14}$N
enhancements and $^{12}$C depletions are observed in globular
clusters, as seen in Figures~\ref{nfe_bulge} and \ref{nc_bulge}.

The above simple CN-cycle mixing scenario cannot account for the
more extreme abundances in our sample.  Most models proposed to
account for the existence of such stars in GCs contend that they
result from chemical evolution within the globular clusters themselves,
as discussed in Section~\ref{intro}. In apparent support to that
scenario, Figure~\ref{nc_bulge}b indicates that the GC sequences
at high [N/Fe] tend to have, on average, lower [C/Fe] than those
at normal [N/Fe], which is a manifestation of the well-documented
fact that SG stars are both enhanced in N and diminished in
C relative to FGs.  Yet, the sample of high [N/Fe] outliers
in Figure~\ref{nfe_bulge} contains stars with [C/Fe] as high as
$\sim$ +0.3.  Because such stars are not typically found in GCs,
and because we want to avoid contamination by objects such as CH
stars \citep[e.g.,][]{ka15,mw90}, we restrict our sample to stars
with [C/Fe] $<$ +0.15, leaving us with a total of 58 N-rich stars.
This additional selection criterion has no impact on the conclusions
presented in this paper.

\begin{figure*}
 \centering
 \includegraphics[width=160mm]{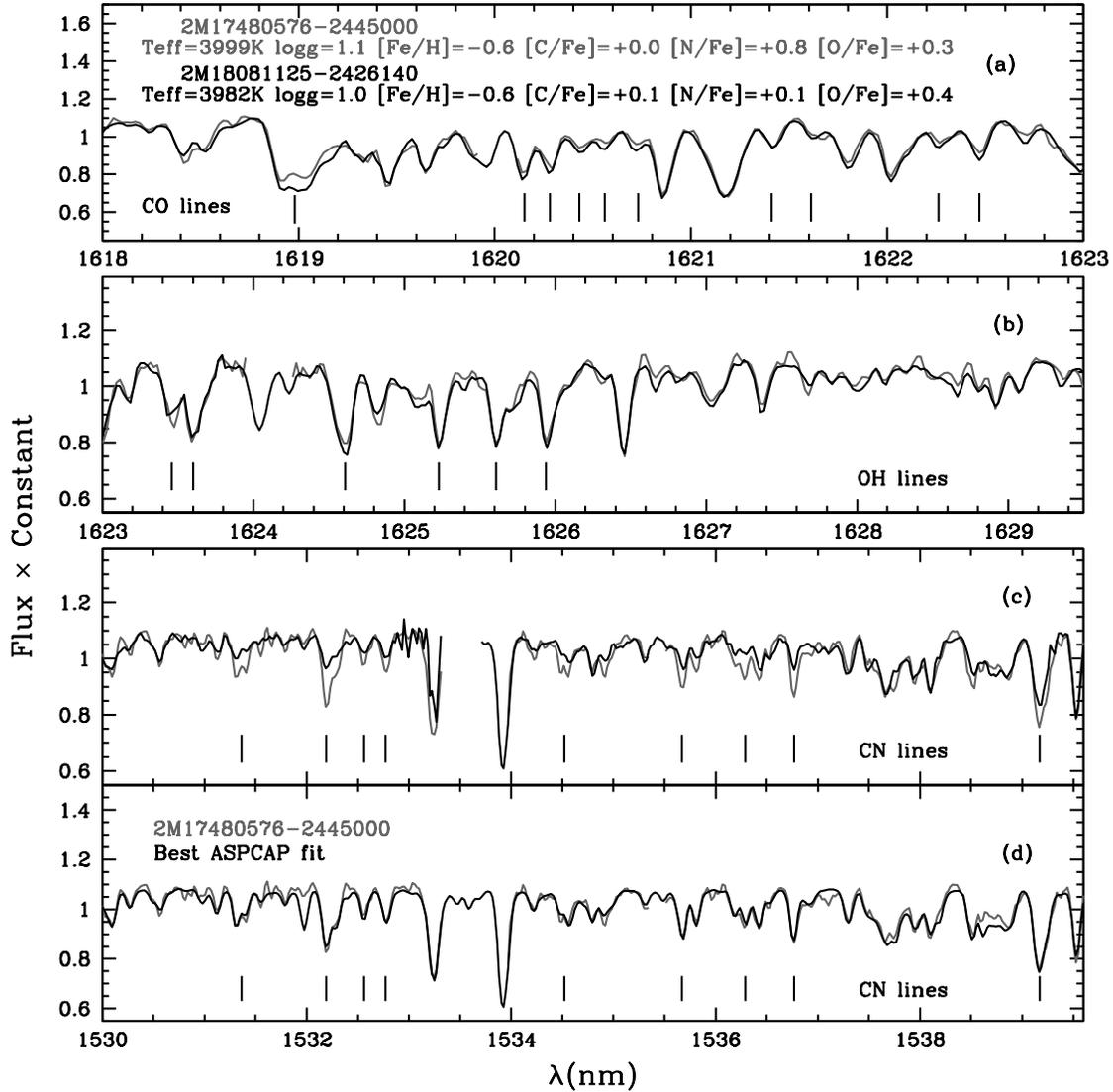}
 \caption{APOGEE spectra and spectral fits. {\it (a)} Comparison between
the spectra of a normal (black) and a N-rich star (gray), with similar
stellar parameters and similar abundances of carbon and oxygen, as
labeled.  The vertical ticks indicate the positions of CO lines, which are similar
in the spectra of the two stars, with the exception of the lines at
$\sim$~1619~nm, which are contaminated by atomic lines (see text).  
{\it (b)} Spectra of the same stars, now
in a region containing OH lines, which are indicated by tick marks.  OH
lines are also similar in the spectra of the two stars.  {\it (c)}
Same spectra in a wavelength range containing CN lines, indicated by
vertical tick marks.  The N-rich star has much stronger CN lines.  Given the
similarity in stellar parameters and the strengths of CO and OH lines, the
difference in CN lines can only be due to the N-rich star indeed having a
much higher abundance of nitrogen.  {\it (d)}  The spectrum of the N-rich star
from panel {\it (c)} is compared to the best fit from ASPCAP.  The match is
excellent, which lends confidence to the quality of nitrogen abundances
delivered by ASPCAP.  The match to the spectrum of the N-normal star (not
shown) is equally good.  Bad detector pixels, as well as those
characterised by strong airglow residuals, are masked.}
 \label{spectra}
\end{figure*}

Figure~\ref{nfe_bulge}b displays field and GC star data together
on the [N/Fe] vs [Fe/H] plane.  The metal-poor globular cluster
M~13 is included in this plot, so that GCs span as wide a metallicty
range as possible, for a fair comparison with the field sample.  It
is clear that the GC population spans a wide range of [N/Fe] at
fixed [Fe/H], due to a combination of deep mixing and stellar
population complexity.  For each GC the data are distributed along
discrete sets of data points, which are separated from each other
by several tenths of a dex in [N/Fe].  The lowest discrete set,
which is the one best sampled, possesses a total (i.e., ``peak to
peak'') scatter of $\sim$ 0.3 dex in [N/Fe].  The individual groups
correspond to each of the multiple stellar populations present in
the GCs, whereas the [N/Fe] scatter within each group is due to
mixing along the giant branch \citep[for more details, see][]{sm03,me15}.
Interestingly, the minimum [N/Fe] for GC stars matches approximately
the values for the field population at same [Fe/H].  In other words,
the [N/Fe] vs [Fe/H] relation of the field population lies along
the lower envelope of the [N/Fe] distribution in the GC samples.
This result is consistent with findings by other studies, which
have characterised FG stars in GCs as having the same chemistry
as field stars of same metallicity \citep[e.g.,][]{ca09}.
The scatter of the field population in [N/Fe] is similar to that
of the GC populations with lowest [N/Fe] (FG stars), suggesting
that the {\it thickness} of the stellar sequences at constant [Fe/H]
is due to deep mixing, at least in the low metallicity regime.


The similarity between N-rich stars and SG stars from GCs
can be further tested through examination of other abundance
(anti-)correlations typically found in GCs.  The most popular is
the anti-correlation between the abundances of Na and O, but
unfortunately sodium abundances in DR12 rely on a NaI line at
$\lambda$~1639.333~nm that is too weak in the spectra of stars with
the typical $T_{\rm eff}$ and metallicity of the N-rich sample for
reliable abundances to be derived by ASPCAP.  Examination of spectral
fits for a handful of cool stars in the metal-rich end of our sample
suggests the sodium abundances to be very high in N-rich stars and
correspondingly lower in N-normal stars.  Further work will be
required to confirm this tentative result.

Aluminium is another element known to present strong variations in
Galactic GCs \citep[e.g.,][]{kr97,gr04,gr12}, whose abundances
should in principle be correlated with those of nitrogen
and anti-correlated with those of carbon.  Fortunately, Al
lines are present in the APOGEE spectral region, so that ASPCAP
abundances are reliable throughout the range of stellar
parameters.  Figure~\ref{aln_bulge}a displays the same data and
symbols plotted in Figure~\ref{nc_bulge}a in the [Al/Fe]-[N/Fe]
plane.  The bulk of the field population, located at [N/Fe] $\simless$
+0.5, shows a slight anti-correlation between the two abundance
ratios, whereas the N-rich stars display a positive correlation.
These trends can be understood by examination of Figure~\ref{aln_bulge}b,
where a sub-sample including only stars with [Fe/H]$<$--0.6 is
compared with APOGEE data for GCs in the same metallicity range.
The SG stars in GCs occupy the same locus as N-rich stars,
reinforcing the similarity between the two populations.  Moreover,
FG stars from GCs occupy the same locus as the general bulge
field, displaying a similar anti-correlation between the two abundance
ratios.


\subsection{Spectra and spectral fits}

Elemental abundances are the foundation upon which the results
presented and discussed in this paper are built, so it is fitting
that we provide evidence in support of the numbers reported by ASPCAP.
A brief examination of typical APOGEE spectra for N-rich and N-normal
stars reassures us of the existence of a real chemical peculiarity
in the former.  Spectral comparisons are presented
in Figure~\ref{spectra}a-c, where the spectra of a N-rich and a
normal star are compared in relevant wavelength intervals.  


Nitrogen abundances in APOGEE spectra are determined solely from
the strengths of CN lines, which in turn are sensitive to other
parameters, chiefly $T_{\rm eff}$, $\log g$, and the abundances of
carbon and oxygen.  The dependence of CN line strength on $T_{\rm
eff}$ and $\log g$ is due to a combination of the well-documented
impact of temperature and pressure on molecular dissociation
equlibrium \citep[e.g.,][]{ru34,ts73} and the ratio between continuum
and molecular line opacity \citep[e.g.,][]{bt91}.  The abundances
of C and O affect CN lines due to their impact on the concentration
of the CN molecule in the stellar atmosphere, via molecular
dissociation equilibrium \citep[e.g.,][]{ru34,ts73}.  To simplify
matters, we control for these parameters, by choosing two stars
with nearly identical $T_{\rm eff}$, $\log g$, [C/Fe], and [O/Fe],
but vastly different [N/Fe], for our comparison.  We stress
that the ``normal'' comparison star is chosen not to be a representative
of a FG counterpart to the N-rich star, but rather to have
stellar parameters and abundances that are identical to that of the
N-rich stars, so as to highlight the impact of nitrogen abundance
variations in the stellar spectrum. The stars selected are
2M17480576-2445000 (N-rich) and 2M18081125-2426140 (N-normal); their
stellar parameters and chemical abundances are displayed in
Figure~\ref{spectra}a.  The spectra of these two stars are compared
in a wavelength range containing several CO lines \citep{me15},
some of which are indicated by vertical tick marks.  One can
immediately conclude that the CO lines are similar between the two
spectra.  In fact, the N-normal star has slightly stronger CO lines,
which is possibly due to a combination of a slightly higher carbon
abundance and slightly lower surface gravity.  The strong bandhead
at $\lambda\sim1619$~nm shows a particularly larger difference,
which is possibly due to contamination by lines due to Si, V, and
Sc.  In Figure~\ref{spectra}b spectra for the same stars are now
compared in a region containing several OH lines and, again, line
strengths are similar in the spectra of the two stars.  The combination
of these two empirical results, in view of the fact that the two
stars have nearly the same atmospheric parameters, means that they
must have similar abundances of carbon and oxygen, as indicated by
the ASPCAP results.  We now turn to Figure~\ref{spectra}c, where
the spectra are compared in a region containing CN lines, again
indicated by vertical tick marks.  The N-rich star has remarkably
stronger CN lines which, in view of the similarity between the two
stars in all the other relevant parameters, can only mean that it
has much higher nitrogen abundance.  ASPCAP tells us that [N/Fe]
in the N-rich star is higher than in the N-normal star by 0.7 dex.
To first order, the quality of the ASPCAP result is verified by
comparison between the observed spectrum and the best ASPCAP fit
(Figure~\ref{spectra}d), where it can be seen that the CN lines
(and indeed most of the spectrum) are well reproduced.  The quality
of the ASPCAP fits to the spectra of N-normal stars can be verified
in other APOGEE publications ~\citep[e.g.,][]{ho15,gp15a}.  We
conclude that the nitrogen abundance differences, which are the
basis for our identification of a new stellar population in the
inner Galaxy, are highly reliable.


\subsection{Evolutionary stage of the N-rich stars} \label{agbs}

Post main-sequence evolution is known to affect the surface abundances
of giant stars during the RGB and AGB evolutionary stages, in ways
that resemble those observed in the N-rich sample
\citep[e.g.,][]{rv81,cl10}.  In particular, the abundance pattern
identified in our sample of N-rich stars is characteristic of the
surfaces of intermediate-mass (3-4 $M_\odot$) AGB stars that have
undergone hot bottom burning.  The presence of such young, moderately
metal-poor stellar populations in the Galactic bulge would have
important implications.  Therefore, the interpretation of our results
depends crucially on establishing the evolutionary stage of the
stars under analysis.  The large luminosities and low temperatures
of AGB stars, combined with the relatively bright APOGEE magnitude
limits in bulge fields \citep{za13} and the focus of our sample on
$T_{\rm eff} < 4500K$ may bias our sample towards a high fraction
of AGB stars.  The possible existence of such a bias is examined
in this section.

\subsubsection{Known AGB stars and colour-magnitude diagram}

We start by searching for known candidate AGB stars in the
N-rich sample.  Inspection of the IDs of AGB stars targeted\footnote{
AGB star candidates were selected from the sample of \cite{sc03},
which is based on H$_2$O and CO absorption, ISOGAL mid-IR excesses,
and light curves.  We refer the reader to that paper for further
details.} in the Galactic centre field (field ID ``GALCEN'' in
the APOGEE data base) showed that none of them is included in the
N-rich sample---in fact, they are almost all too cool ($T_{\rm eff}
< 3500$~K) for ASPCAP to deliver reliable abundances.  The only
exception is star {\tt 2M17451937-2914052}, for which ASPCAP finds
$T_{\rm eff}= 3690\pm91$K, [Fe/H]=$-0.68\pm0.04$, [C/Fe]=$+0.54\pm0.05$,
[N/Fe]=$+0.16\pm0.08$, [Al/Fe]=$-0.10\pm0.10$, which places this
star clearly outside the chemical composition locus occupied by
N-rich stars.

\begin{figure}
 \centering
 \includegraphics[width=84mm]{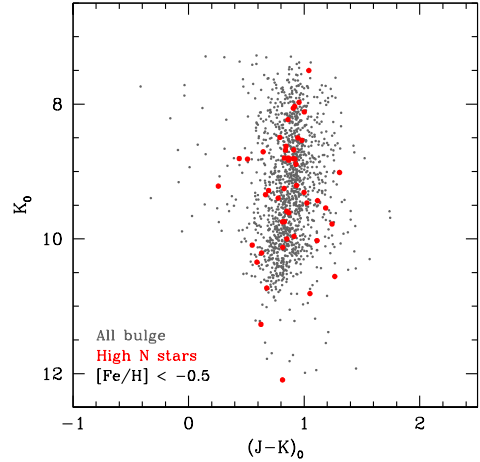}
 \caption{The dereddened 2MASS colour-magnitude diagram for the sample
 studied in this paper.  ``Normal'' stars are plotted using small gray symbols,
 and N-rich stars with larger red symbols.  Only stars with [Fe/H] $<$ --0.5 are
 shown for both samples, to minimise effects due to differences between the
 metallicity distributions of the two samples (the normal stars extend
 to much higher metallicities; see Section~\ref{metdis}).  There are no
 noticeable differences between N-rich and N-normal stars, suggesting that
 the N-rich stars are unlikely to be dominated by AGB stars.}
 \label{cmd2mass}
\end{figure}

The next obvious way of checking for the presence of an AGB bias
is by comparing the distribution of N-rich and N-normal stars in
the colour-magnitude diagram (CMD).  Figure~\ref{cmd2mass} displays
the N-rich and N-normal samples in the dereddened 2MASS \citep{sk06}
CMD.  Dereddening was performed using $A_K$ from DR12 (adopting the
{\tt A\_K\_TARG} parameter), which was inferred through the RJCE
method \citep{maj11}, and adopting the extinction law from \cite{in05}.
The sample plotted is limited to [Fe/H] $<$ --0.5, to minimise
differences between N-normal and N-rich samples that are purely due
to differences in their metalliticy distributions (see discussion
in Section~\ref{metdis}).  One can immediately notice that the two
sub-samples occupy the same locus in the CMD, suggesting that the
N-rich sample is not biased towards AGB stars relative to the
remainder of the field sample.  In other words, the AGB/RGB ratio
of the N-rich sample is likely to be the same as that of the rest
of the field sample.  The same conclusion is drawn from comparison
of the two samples in a (reddening and distance independent) $T_{\rm
eff}$ vs $\log g$ diagram (not shown).  A more quantitative
assessment can be made by selecting N-rich and N-normal stars within
a narrow range of metallicities and compare the differences in colour
between the two samples for the same magnitude, where AGB and RGB
stars can differ in $J-K$ by as much as 0.05 mag \citep[e.g.,][]{gi00}.
We proceeded by fitting fiducials to the two samples in the CMD and
comparing the colours of the fiducials for a given magnitude.  For
instance, at H = 9, we obtain $J-K = 0.91 \pm 0.04$ and $0.91 \pm
0.06$ for N-rich and N-normal, respectively, and at H=10 we obtain
$J-K = 0.85$ for both samples, with same uncertainties.  In conclusion,
we find no difference between N-rich stars and the rest of the field
sample in the dereddened 2MASS CMD, which is consistent with no
substantial difference in AGB contribution to the N-rich sample and
the rest of the field.  

\subsubsection{Infrared excess}

\begin{figure}
 \centering
 \includegraphics[width=84mm]{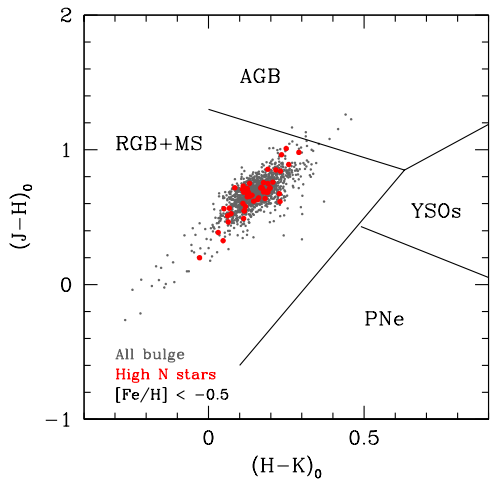}
 \caption{Sample stars in the 2MASS dereddened colour-colour plot,
 adopting same symbols as Figure~\ref{cmd2mass}.  The colour-colour plane is
 divided into sections populated by various types of NIR-bright objects,
 following the study by \protect\cite{ga97}, as indicated by the labels.  
 The vast majority of the sample occupies the RGB+MS region of the diagram, 
 which shows that there
 are very few dusty AGBs in the sample.  In particular, none of the N-rich
 stars occupies the AGB part of the diagram.}
 \label{cc2mass}
\end{figure}

The presence of AGB stars in our sample can be further assessed by
detection of photometric signatures of the presence of circumstellar
dust.  Vigorous mass loss during late stages of AGB evolution is
responsible for the formation of dusty envelopes \citep[e.g.,][]{hab96},
which manifest themselves through excess radiation at long wavelengths
\citep[e.g.,][]{ga97}.  In Figure~\ref{cc2mass} N-normal and N-rich
stars are shown in a 2MASS dereddened colour-colour diagram.  The
loci occupied by RGB and main sequence stars, AGB stars, young
stellar objects (YSOs), and planetary nebulae (PNe), according to
the study by \cite{ga97}, are separated by lines and indicated by
labels.\footnote{The loci of the various object types in the
colour-colour diagram were defined by \cite{ga97} on the basis of
photometry on the Koornneef system \citep{ko83}.  Considering small
zero point differences between different photometric systems and
an error of 5-10\% in photometry of bright stars, the mismatch
between the loci in the two systems should be at most 0.1 mag, which
does not affect our conclusions.} Because of the presence of
circumstellar dust, AGB stars occupy a locus towards colours redder
than those of RGB and MS stars in this diagram.  From the distribution
of the data points, one can see that the vast majority of the 
sample is located in the RGB+MS sequence, with only about 1\% of
all the stars positioned in the AGB area of the plot.  In particular,
all N-rich stars inhabit the RGB+MS part of the diagram, which
provides further evidence for the absence of an important contribution
of AGB stars to our N-rich sample.

\begin{figure}
 \centering
 \includegraphics[width=84mm]{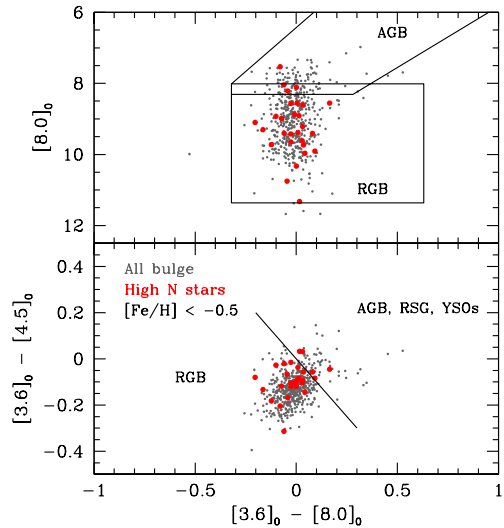}
 \caption{Using dereddened {\it Spitzer} IRAC photometry to determine
  the evolutionary stage of our sample.  {\it Top panel}: The loci of
  N-normal and N-rich stars in the CMD is displayed.  The areas of the
  diagram occupied by AGB and RGB stars are shown, suggesting that both the
  N-normal and N-rich samples are dominated by RGB stars. {\it Bottom
  panel:} Same sample as in the top panel, now in a colour-colour diagram.  
  The solid line shows the frontier between the loci of RGB and other stellar 
  types.  Again, the evidence suggests that our sample is overwhelmingly 
  dominated by RGB stars.}
 \label{iracphot}
\end{figure}

Figure~\ref{iracphot} shows our sample stars on colour-magnitude
and colour-colour diagrams based on {\it Spitzer} IRAC dereddened
photometry \cite[again using the extinction law by][]{in05}.  Mid-IR
colours are particularly sensitive to the presence of dust, so these
diagrams help spotting AGB stars with dusty envelopes.  On the top
panel, N-normal and N-rich stars are contrasted on a CMD, where the
loci of RGB and AGB stars at a distance of 8 kpc, according to
\cite{ra08} are indicated.  As in previous cases, the data strongly
suggest of a very small contribution by AGB stars to our sample.
In the bottom panel, stars are displayed on an IRAC dereddened
colour-colour diagram, where a line separates the locus of RGB from
that of redder YSOs, red supergiants, and AGB stars.  The loci were
established from visual expection of Figure 1 (bottom panel) of
\cite{ma07}, who studied IRAC photometry for a sample of Galactic
AGB stars.  Again in this case, the vast majority of our sample
stars is located safely outside the AGB region of the diagram, and
within the locus commonly occupied by RGB stars.

\subsubsection{Variability}

Asymptotic giant branch stars undergo thermal pulsations
\citep[e.g.,][]{ir83}, which manifest themselves observationally
in the form of moderate to high amplitude, long-period magnitude
variations.  Stellar types commonly associated with the AGB phase
are Mira-type variables, long-period variables, semiregular variables,
and the heavily obscured OH/IR stars \citep[e.g.,][]{gc09,ji06}.

\begin{figure}
 \centering
 \includegraphics[scale=0.5]{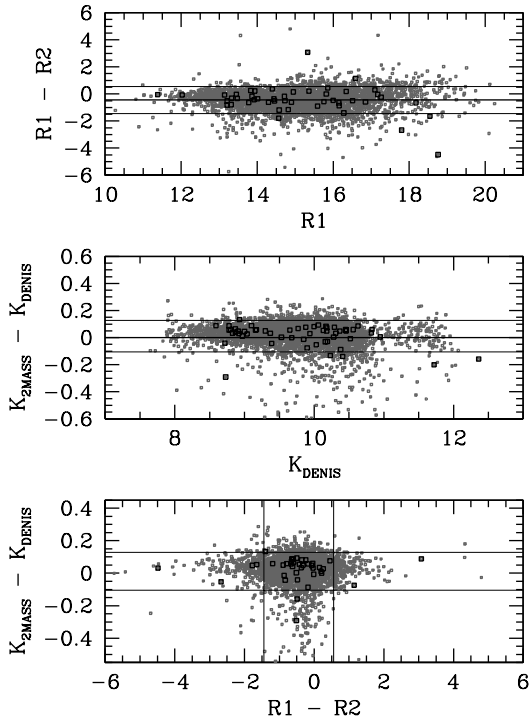} \caption{Assessment
 of the amoung of variability in the N-rich sample.
{\it Top panel:} Variation in R magnitude between two epochs in the
USNO-B catalog, as a function of magnitude in the first epoch.
Small gray and solid black dots normal and N-rich stars, respectively.
The thick horizontal line marks the position of the mean difference,
and the two thin lines mark $\pm2\sigma$ departures from the mean,
where $\sigma$ is the photometric precision.  The mean difference
and standard deviation are the same for the N-rich and normal sample.
Moreover, about 90\% of the points in both samples are within
$2\sigma$ of the mean, thus having no evidence for variability.
{\it Middle panel:} Same as above, for the difference between K-band
magnitudes in 2MASS and DENIS, which have negligible zero-point
differences.  Conclusions are the same as in top panel.  {\it Bottom
panel:}  Comparison between magnitude variations in optical and NIR
from panels above.  The thin lines again indicate $2\sigma$ off the
mean values, where $\sigma$ is again photometric precision.  There
is no correlation between magnitude variations.  Stars for which
there is an indication of variability in one band are consistent
with no variation in the other band, suggesting that photometric
errors may be responsible for strong variations observed.  }
 \label{deltamags}
\end{figure}

To establish the occurrence of variability, one ideally needs
multi-epoch observations, preferably performed within the same
photometric system, to minimise confusion due to zero-point
differences.  These are difficult to obtain because of the long
periods associated with these variations, typically of the order
of $10^2-10^3$ days \citep{gc09}.  In light of these requirements,
probably the best source for photometric variability information
on our sample stars would be long-term monitoring photometric surveys
over large areas of the Galactic bulge, such as OGLE \citep{ud97,so13}
in the optical, and VVV \citep{sa12,ca13} in the NIR.  Unfortunately,
the overlap between the APOGEE and OGLE footprints is very small,
so that only 4 N-rich stars were observed by OGLE.  None of these
stars are included in the OGLE catalogs of long or double period
variables.  The overlap with the VVV footprint is much larger, but
due to the relatively bright limit of the APOGEE bulge sample
(H$<$11.2), saturation is a problem for VVV data \cite[see, e.g.,][for
details]{ca13}.  We first examine magnitude differences between the
two epochs included in the USNO-B catalog \citep{mo03}, which
contains 4,296 stars in common with our bulge sample, of which 55
are N-rich stars.  The data are shown on the top panel of
Figure~\ref{deltamags}, where R-band magnitude variations are plotted
as a function of first-epoch magnitudes.  The thin lines indicate
$2\sigma$ departures from the mean difference (thick line), where
$\sigma$ is the photometric precision ($\sim 0.5$ mag).  Most stars
are consistent with no variability.  About 11\% of the N-rich stars
varied by more than twice the photometric precision, whereas when
the entire sample is considered, 13\% of the stars varied by more
than $2\sigma$.  We find, for the whole sample, $\langle\Delta {\rm
R1}\rangle = -0.4 \pm 0.9$ and for N-rich stars we find $\langle\Delta
{\rm R1}\rangle = -0.4 \pm 1.0$.  Combined, these numbers suggest
that the N-rich stars do not include a higher fraction of variable
stars than the rest of the sample.  This result is insensitive to
assumptions on the photometric precision of the UNSO-B catalog.
Moreover, comparison of DENIS and 2MASS data (below) suggests in
fact that the variability information as inferred from this analysis
of USNO data is actually questionable.

Identification of variable stars in the USNO-B catalog is made
difficult by its relatively low photometric precision.  Therefore,
we investigate more precise photometry from NIR catalogs, which
also have well determined photometric zero points.  Comparing 2MASS
\citep{sk06} with UKIDSS \citep{la07} data, we found 36 N-rich stars
are present in both catalogs out of a total common sample of 360
stars.  Unfortunately, saturation of UKIDSS data is a problem for
stars brighter than $K_{2MASS} \sim 10.5$, but the few stars fainter
than that limit show no evidence for variations between the two
catalogs.

We next examine possible variations between DENIS \citep{ep97} and 2MASS magnitudes,
noting that \cite{cg03} showed zero point differences between the
two photometric systems to be not significant.  A total of 4,692
stars from our sample are included in both catalogs, with 58
of them being N-rich.  In Figure~\ref{deltamags}, middle panel, the
difference between K-band magnitudes in the 2MASS and DENIS systems
is plotted as a function magnitude.  Symbols and horizontal lines
have the same meaning as in the top panel, adopting a photometric
precision of 0.03 and 0.05 mag for 2MASS and DENIS, respectively
\citep{sk06,ci00}.  The result of this comparison is consistent
with what we found from analysis of the USNO-B data, with the vast
majority of the stars being consistent with no variation.  About
$\sim$15\% of the stars in the sample have variations larger
than $2\sigma$, and in the N-rich sample, the number is $\sim$12\%.

It is possible that some of the stars below the variability detection
threshold of this comparison are small amplitude red giant variables,
such as those identified in the OGLE survey \citep{wr04,so13}.  This
in fact may partly explain why the scatter in the residuals
($\sim$0.1--0.14 mag) is larger than expected on the basis of
photometric errors alone ($\sim$0.06 mag).  Most importantly, if
the incidence of AGB stars was higher in the N-rich sample, we would
expect magnitude differences for those stars to present a larger
scatter than for the rest of the sample.  However, in agreement
with USNO-B data, this is not what we find.  In fact, the standard
deviation of the N-rich sample is slightly smaller in the N-rich
sample ($\langle\Delta {\rm K}\rangle = 0.01 \pm 0.11$) than in
the rest of the sample ($\langle\Delta {\rm K}\rangle = 0.01 \pm
0.14$), which taken at face value suggests that the incidence of
variable stars in the N-rich sample is, if anything, smaller than in
the rest of the sample.

In conclusion, analysis of two-epoch observations in both the optical
and NIR consistently indicate that the fraction of variable stars
in our sample is of the order of $\sim$10\%.  However, interestingly,
the bottom panel of Figure~\ref{deltamags} shows that there is no
correlation between magnitude variations in the USNO and 2MASS/DENIS
catalogs, particularly in the N-rich sample.  It is noteworthy that
the stars for which variation is largest in one band, in the other
band fall within the area of the graph consistent with no variation.
In other words, they are confined to the cross-shaped locus defined
by the $\pm 2\sigma$ limits in both bands.  In other words, stars
that look variable in NIR do not in optical, and vice versa.  This
may be partly due to the fact that observations were collected at
different epochs, but in this case one would expect that at least
{\it some} of the largest deviants would fall outside the cross-shaped
locus.  Of relevance to this discussion is the fact that magnitude
errors for the most deviant points in the USNO-B sample are
particularly large, even though they do not entirely account for
the discrepancies.  All in all, the bottom panel of Figure~\ref{deltamags}
suggests that the fraction of variables in the N-rich sample may
be indeed smaller than suggested by USNO and 2MASS-DENIS data
considered in isolation.

We conclude that there is little evidence for a large
contribution of AGB stars to our sample.  Considering uncertainties
in photometry and in the locus occupied by the various stellar types
in the 2MASS and IRAC CMDs and colour-colour plots, it seems safe
to conclude that our sample stars are predominantly composed of
first-ascent RGB stars, with at most a small contribution by AGB
stars.  It is difficult to provide a solid estimate of the ratio
between the number of AGB and RGB stars in our sample, without a
more detailed analysis of the AGB candidates.  Examination of the
data presented in this Section suggests that the ratio is probably
no larger than $\sim$ 5-10\%, being thus consistent with data from
other old metal-poor stellar populations \citep[see, e.g.,][for a
discussion]{gi10}.  This topic will be the subject of further
investigation in a future publication (Zamora et al., 2016, in
prep.).

\subsection{Incidence of Mass Transfer Binaries in N-rich Sample}
\label{binaries}

Stars with N and Al overabundances can also originate through a
binary mass-transfer channel.  Intermediate-mass stars
(M$\sim$3-8\,M$_{\odot}$) undergo hot bottom burning during their
AGB phase, producing large amounts of N and Al \citep[e.g.,][]{ve13}.
When these objects are members of a binary with an appropriate
semi-axis size, mass transfer takes place and the low-mass companion
atmosphere is enriched with the products of AGB nucleosynthesis.
The donor star evolves away from the AGB phase and eventually becomes
a faint white dwarf, while the companion retains their chemical
signature.

Establishing the presence of radial velocity variations among the
N-rich stars would be the most natural course of action towards
estimating the fraction of such objects formed through the binary
channel.  However, most of the objects in our sample were observed
just once and those with multiple observations have a relatively
short ($\leq$ 6 months) baseline, which makes possible detection
of only a small fraction of such binaries.  Observations of CH
stars, which are objects of similar nature (see below for details),
are typically of the order of several years \citep[see, e.g.,][]{lu05a}.

To determine the expected number of mass-transfer binaries in a
given population from first principles one needs to know a number
of properties of the underlying stellar population, such as the
initial mass function, the binary fraction as well as binary period,
eccentricity, and mass ratio distributions.  Systematic studies of
orbital properties of binary stars have been generally limited to
the solar neighbourhood, and mostly deal with solar type
\citep[e.g.,][]{dm91} or M-dwarf stars \citep[see, e.g.,][]{aw04,fm92}.
Nevertheless, reasonable assumptions can be made for these quantities.
However, another critical ingredient is the range of periods (or
semi-axes) within which mass transfer takes place effectively, which
is highly uncertain, due to the current incomplete understanding
of mass transfer during the common envelope phase and the treatment
of angular momentum loss \citep[for a discussion of the
theory and examples of its applications, see][respectively]{ab13,ab15}.


Alternatively, one can estimate the expected fraction of
N-rich stars which result from transfer binaries in a more empirical
way, by using the observed number of CH stars in the whole
sample. Classical CH-stars, like Ba and CEMP-s stars owe their
peculiar composition to mass transfer from a relatively low-mass
(M$\sim$1.5-4\,M$_{\odot}$) companion \citep[see, e.g.,][]{mw90,lu05a,st14}.
The mass range for the donor star is determined by the minimum mass
for the third dredge-up (and hence for becoming a CH star) and
by the onset of effective hot bottom burning, which burns C into N
quite effectively.

Under the assumption that the binary incidence and the distributions
of orbital period, mass ratio, and eccentricity are not dependent
on the mass of the primary (which is quite reasonable in the mass
range under discussion) and that the same mass-transfer physics
applies, the ratio between the expected incidence of CH stars and
N-rich stars should be equal to the number ratio of donor stars in
a given population. These numbers can be easily estimated by assuming
that companions to CH stars and N-rich stars had initial masses in the
1.5-3\,M$_{\odot}$ and 3-8\,M$_{\odot}$ range, respectively.  This
ratio is $\sim$ 0.5, being rather insensitive to whether one picks
a \cite{sa55}, \cite{kr01} or \cite{ch03} IMF.  This is likely an
overestimate: while all stars in the $\sim$3-8\,M$_{\odot}$ range
undergo hot bottom burning and hence become enriched in nitrogen,
the Mg-Al cycle is activated at $T\sim50\,MK$, and considerable
Al production happens only for stars with masses within the higher
end of the above range \citep[with the lower cutoff depending on
metallicity see, e.g.,][]{ve13}. Using this ratio, we can determine,
on the basis of the number of {\it bona fide} CH stars, the expected
number of {\it bona fide} N-rich stars of binary origin.

In order to count the number of CH-star candidates in our sample,
a lower limit in [C/Fe] must be defined.  A commonly adopted number
for metallicity within $-1 \simless$~[Fe/H]~$\simless 0$ is [C/Fe]=+0.3
\citep[see e.g.,][]{lu05b}.  However, [C/Fe] varies slightly with
metallicity within the sample, so that a sample of CH-star candidates
defined on the basis of a constant lower limit would be biased
towards stars with higher metallicity.  Therefore, CH-star candidates
are selected in a manner analogous to how the N-rich sample itself
was defined, by fitting a high order polynomial to the run of [C/Fe]
with [Fe/H] and taking stars deviating from the fit by $+4\sigma$.
The CH-star candidate sample defined in this way has a metallicity
distribution in acceptable agreement with that of the N-rich stars
(Section~\ref{metdis}), which is consistent with the two samples
belonging to the same underlying stellar population. The CH-star
candidate sample contains 52 stars. Examining the ASPCAP outputs
for these stars, we find that 23 have one or more of several ASPCAP
quality flags raised, namely, {\tt STARWARN, ROTATIONWARN, CHI2
WARN, COLORTE WARN, and TEFF WARN}.  This indicates, as confirmed
by visual inspection of several examples, that the spectral fits
are of poor quality, or that the stellar parameters are unreliable,
or both, suggesting that the abundances for these stars cannot be
relied on (note that none of those flags were raised for any of the
N-rich stars).  Eliminating these 23 stars from the sample and
recalling that the expected ratio of N-rich to CH stars resulting
from binaries is ∼0.5, our estimate of the number of N-rich stars
owing their atmospheric composition to binary mass transfer would
be $\sim$ 15, or approximately a quarter of the sample identified
in previous Sections.

We emphasise that this number is an upper limit, based on adopting
a wide mass range for the production of Al in intermediate-mass
stars and a low cutoff for considering a star C-rich. In fact, we
argue that the real number must be smaller, for the following
reasons. First and foremost, one would expect to find a population
of N-rich stars in the Galactic disk if the phenomenon had an
important contribution from mass transfer binaries. There are
$\sim$~95 CH-star candidates, defined in the same way as above, in
a disk sample defined by $|b| < 20^\circ, 20 < l < 340^\circ$, and
the same atmospheric parameters as the bulge sample. Of those, 46
have ASPCAP quality flags raised, leaving us with a sample of 49
CH-star candidates.  Following the same reasoning as above, one
would expect to find ∼ 25 N-rich stars in the disk field, and instead
not a single one is found. In fact, the few N-rich stars identified
within the thusly defined disk belong to the low latitude globular
clusters M 71 and NGC 6760. One might argue that, because the disk
has a higher overall metallicity than the N-rich sample, the
production of N-rich stars is inhibited, because models suggest
that intermediate-mass AGB stars produce less nitrogen at higher
metallicity \citep{ve13}.  However, looking at the problem in a
different way, we can examine the frequency of metal-poor N-rich
stars in the bulge and ask how many such stars we would expect in
the disk. At [Fe/H]~$<-1.0$, there are 15 N-rich stars out of a
sample of 220 bulge stars in the same metallicity range and with
same constraints on [C/Fe]. In the disk, the number of stars in the
same locus of parameter space is 100, so that one would expect about
$7 \pm 3$ N-rich moderately metal-poor disk stars to be discovered
by APOGEE, in disagreement with the absence of any such stars in
our sample.

Therefore, we conclude that, to the best of our knowledge, the
contamination of the N-rich star sample by the remnants of mass
transfer binaries can amount to as much as 25\%, although the fraction
is likely to be smaller.

\bigskip

We summarise the content of Sections~\ref{agbs} and \ref{binaries} by
concluding that there is no evidence for a high incidence of either
AGB stars or mass-transfer binaries in our sample, so that the
abundance pattern observed in most of the sample cannot be explained
by those phenomena.  We thus conclude that we have identified a
stellar population in the inner Galaxy with a chemical composition
akin to that of SG stars from globular clusters.  The
implications of this result are discussed in Section~\ref{discussion}.

\subsection{Metallicity Distribution} \label{metdis}

\begin{figure}
 \centering
 \includegraphics[scale=0.5]{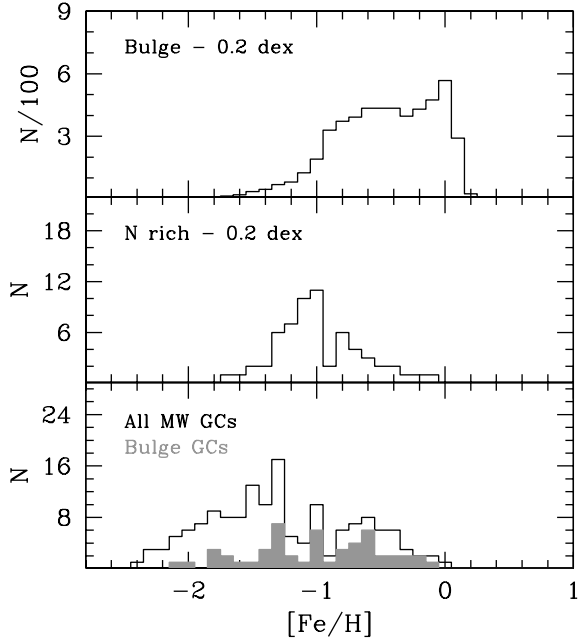}
\caption{Metallicity distribution functions (MDFs) for the bulge
field (top panel), N-rich stars (middle panel) and the Galactic
globular clusters (bottom panel).  The hatched gray histogram in
the bottom panel shows the MDF obtained when only bulge GCs are
considered.  The MDFs of the bulge field and N-rich stars have
significantly different shapes, so that it is difficult to conceive
of a scenario where dissolved GCs contribute significant amounts
of stellar mass to the Galactic bulge.  The MDF of N-rich stars is
also quite different from that of the Galactic GCs either considering
the entire GC system (open histogram) or only those contained within
the volume defined in Section~\ref{data} (hatched gray), making it
difficult to conceive of a single mechanism connecting the population
of dissolved GCs with the existing ones.  See text for details.
}
 \label{mdfs}
\end{figure}

In order to characterise the newly discovered stellar population,
we examine its metallicity distribution function (MDF) and, in
Section~\ref{discussion}, contrast it with those of the Galactic
bulge and globular clusters.  Figure~\ref{mdfs} shows the MDFs for
these three samples.  The top panel shows the MDF of the entire
population defined in Section~\ref{data}, whereas the middle panel
shows that of the N-rich stars.  The bottom panel shows the MDF of
Galactic GCs included in the 2010 version of the Harris catalog
\citep{ha96}.  The metallicities of both APOGEE samples
are corrected by --0.2 dex, to bring the APOGEE metallicity scale
for [Fe/H]$\simless$--0.4 into agreement with the literature on
abundance studies in the optical \citep[see][for details]{ho15}.
A constant correction is adopted for simplicity, even though
it is only good for stars within the above metallicity range.  As
one can attest from inspection of Figure 6 of \cite{ho15}, raw
[Fe/H] values provided by ASPCAP are in good agreement with the
literature for metal-rich clusters, so that the corrected values
for metal-rich are too low by 0.2 dex.  Since the number of N-rich
stars in this metallicity regime is negligible, this small inaccuracy
does not affect our results or conclusions.

\subsubsection{The bulge MDF} \label{bulgemdf}

A detailed examination of the MDF in the top panel of Figure~\ref{mdfs}
is beyond the scope of this paper, and for that we refer the reader
to \cite{gp15b}.  For our purposes, we simply state that the global
bulge MDF is in good agreement with those by \cite{ra14} and
\cite{ne13a}, which are based on Gaia-ESO and ARGOS data, respectively.
In particular, stars with [Fe/H] $<$ --1 make up 5.8 and 4.6\% of
the samples by \cite{ra14} and \cite{ne13a}, respectively, whereas
they make up 228 out of our sample of 5,140 stars, or 4.4 
$\pm$ 0.4\%, which is in formal agreement with the \cite{ne13a}
MDF.

To err on the side of caution, we checked for the presence
of a bias against metal-poor giants that could have been introduced
by the $T_{\rm eff}<4500K$ limit in our sample.  We found that
extending the sample by including giants as warm as $T_{\rm eff}=6500K$
added 18 stars to the sample, increasing only marginally the
percentage of stars with [Fe/H]$<$--1, to 4.7 $\pm$ 0.4\%.  Examining
in detail the $T_{\rm eff}$ distribution of the stars that are
excluded by the $T_{\rm eff}<4500$K cut, we find that all stars
have $T_{\rm eff}<5100$K.  The absence of warmer stars is due to
the relatively shallow (H=11) APOGEE magnitude limit in bulge fields,
which restricts the sample to cool evolved giants.  Regarding their
metallicity distribution, the stars range between --2$<$[Fe/H]$<$+0.0,
with a mean value of [Fe/H]=--0.94.  On the metal-poor end, 16 stars
have [Fe/H]$<$--1 to be contrasted with at sample of 220 stars in
the same metallicity range and $3500<T_{\rm eff}<4500$K.  Most
importantly, only at most 5/18 of those stars can be classified as
N-rich following the definition provided in Section~\ref{nrich}.
Of those, 2 have [Fe/H]$>$--1, 2 are within the --1.5$<$[Fe/H]$<$--1
interval, and 1 has [Fe/H]$<$--1.5.  Such a small sample of warm
N-rich stars has obviously no impact on the MDF of the N-rich
population.  It is thus safe to conclude that our sample is free
of any important metallicity bias on the metal-poor end.  In closing,
we note that the N-rich/N-normal ratio within this warm sub-sample
is 5/18, thus substantially higher than in the rest of the metal-poor
sample.  Given the considerable uncertainties in [N/Fe] at such low
metallicity and relatively high $T_{\rm eff}$, we ascribe little
significance to this result, while deeming it worthy of a closer
examination in the near future.  




\section{Discussion}  \label{discussion}

The findings discussed in the previous section tantalizingly suggest
that a population of stars with globular cluster origin has been
identified in the inner Galaxy.  In line with early theoretical
predictions \citep[][see also Gnedin et al.\ 2014]{tr75}, we
hypothesise that these stars result from the destruction of
pre-existing Galactic globular clusters.  Interestingly, \cite{bk15}
proposed that millisecond pulsars, resulting from the destruction
of GCs, can account for gamma ray detections by the {\it Fermi}
satellite towards the central regions of the Galaxy.  This has
important implications, as \cite{bk15} show that their model does
a better job of matching the data than models invoking annihilating
dark matter \citep{hg11}.  Assuming this hypothesis is correct, our
discovery may also have interesting repercussions for the current
understanding of the formation and evolution of the Galactic globular
cluster system, the presence of multiple stellar populations
in globular clusters, and possibly also the formation of the bulge
itself.  In this section we discuss some of these ramifications.
We conduct our discussion in Sections~\ref{lowmet} through \ref{nvsgcs}
within the framework of a GC origin for the newly found stellar
population.  In Section ~\ref{ETGs}, we speculate on a possible
connection between N-rich stars and stellar populations inhabiting
the cores of Andromeda and early-type galaxies.  Finally, in
Section~\ref{origin} we discuss the origin of the new stellar
population, conjecturing also possible scenarios beyond a pure GC
origin.

\subsection{MDF constraints on the FG/SG ratio} \label{lowmet}

We begin by exploring the observed MDF in order to place
constraints on the ratio between the numbers of stars with N-normal
and N-rich chemical compositions---the FG/SG ratio---in the parent
population of N-rich stars.  Figure~\ref{mdfs} shows that the MDF
of N-rich stars (middle panel) differs quite substantially from
that of the rest of the full sample (top panel).  While the bulge MDF
extends towards very high metallicities, that of N-rich stars peaks
around [Fe/H]=--1, with broad wings towards high and low metallicity.
The apparent dip around [Fe/H]~=~--0.9 is strongly dependent on the
binning adopted, so we assign no significance to it, given the small
numbers per bin.


Assuming that the MDF of N-rich stars reflects precisely that of
the destroyed GCs, one can use the metal-poor tail of the MDFs to
constrain the ratio of N-normal to N-rich stars, and in this way
derive an upper limit on the mass lost by GCs.  There are 15 N-rich
stars with [Fe/H]~$<-1$ out of a sample of 214 stars in the field
within the same metallicity interval and [C/Fe]~$<+0.15$. In other
words, approximately 93\% of the stars in our sample with [Fe/H]~$<-1$
have normal N abundances.  Therefore, if one was to accept
that {\it all} bulge stars with those metallicities originate from
the destruction of GCs---a rather extreme scenario---one would
conclude that 93\% of the stellar mass in those systems was originally
in the form of FG stars. We emphasise that this is an {\it
upper limit}.  The MDF in the top panel of Figure~\ref{mdfs}
shows that the inner Galaxy is dominated by a metal-rich stellar
population which most likely does not result from GC destruction,
seeing as it lacks a counterpart in the N-rich MDF.  It is only
reasonable to assume that some of those field populations not
associated with GC destruction are also present at [Fe/H]$<-1$.
Thus, to have all bulge stars with [Fe/H]$<-1$ come from GC destruction
is most likely an unachievable feat.  Consequently, the fraction
of FG stars in those systems was most likely lower than 93\%.
If we were to assume, for instance, that as much as 1/2 of
the mass in the volume sampled resulted from GC destruction, then
we would conclude that FG/SG $\sim$ 6 for the dissolved GCs.  In
subsequent sections we explore the impact of the acceptable range
of FG/SG ratio values on estimates of the total stellar mass in
destroyed GCs.  This result rules out models requiring FG/SG $\sim
10-100$ to address the mass budget problem.

Most importantly, this result poses a constraint on models that
address the mass budget problem in GC formation by proposing that
the FG/SG ratio was much larger in early GCs than observed
today.
Recently, \cite{lar12,lar14} determined integrated-light metallicities
of GCs in dwarf galaxies, and established that the ratio between
the stellar mass in GCs and the field, at the same metallicity, in
the stellar haloes studied was too high to accommodate the requirements
from selective stellar mass-loss models. That essentially the same
conclusion has been reached from studies of samples in very different
environments and metallicity ranges suggests that this may well be a
general result.

\subsection{Mass in Dissolved GCs} \label{dissolve}

Under the hypothesis that the N-rich stars result from dissolution
of Galactic GCs, it is interesting to use their observed numbers
to estimate the total stellar mass contained in dissolved GCs, to
assess its contribution to the total stellar mass of the inner
Galaxy and also to compare it with the mass contained in surviving
GCs.  However, only one type of GC star can be uniquely discriminated
on the basis of the data under consideration---those belonging to
the SG group.  An estimate of the total stellar mass allocated
in the form of these dissolved clusters obviously requires knowledge
of the contribution by FG stars, which cannot be distinguished
from field populations of the same metallicity, at least not on the
basis of APOGEE data alone.  In the absence of any constraints on
the FG/SG ratio, we make assumptions for two limiting cases
that hopefully bracket the entire range of possibilities.  
The {\it minimal scenario}, assumes that there is no mass
budget problem.  In other words, the FG/SG ratio is exactly
as observed today---about 1/2 \citep{ca09}.  In contrast, in
the {\it maximal scenario}, FG stars completely overwhelm
N-rich stars.  According to that scenario, early GCs were 
10-100 times more massive in the past \citep{bl15,cz15}, with
essentially all the mass lost having been in the form of FGs.
However, as discussed in Section~\ref{lowmet}, the FG/SG ratio
cannot have been higher than $\sim$ 9/1 without violating constraints
from the low-metallicity end of the MDF discussed in the previous
section.

We discuss these scenarios and their consequences in Sections~\ref{minimal}
and \ref{maximal}.

\subsection{The Minimal Scenario}  \label{minimal}

To estimate the mass in dissolved GCs, we first estimate the
fraction of the total stellar mass in the inner Galaxy contributed
by N-rich stars, and following that we determine the contribution
by dissolved GCs assuming FG/SG = 1/2. The total number of N-rich
stars, selected as described in Section~\ref{distrib}, is 58 out
of a total sample of 5,140 stars.  To first order, the ratio
of N-rich to N-normal stars is not biased in any important way by
APOGEE's target selection criteria \citep[see][for details]{za13},
or by the definition of our sample, including its range of
stellar parameters (Section~\ref{data}).  Therefore, we can safely
state that N-rich stars amount to about 1.1\% of the total population
in the inner Galaxy.

Assuming FG/SG = 1/2, we conclude that the contribution
of dissolved GCs to the stellar mass content of the inner Galaxy is
small, at the 1.7\% level.  By further assuming a (unlikely)
scenario where the contribution of mass-transfer binaries to the
N-rich sample is at its maximum 25\% level, we would be led to
conclude that the lower limit for the fractional contribution of
dissolved GCs to the mass of the bulge is 1.3\%.


The total mass of the Galactic bulge is estimated to be
$\sim 2 \times 10^{10}~M_\odot$ \citep{so09}, with a stellar mass
ranging somewhere between 1.25 and 1.6$ \times 10^{10}~M_\odot$
\citep{po15}.  In the minimal scenario, the total mass in stars
resulting from GC destruction would then range between 1.6 and $2.1
\times 10^8~M_\odot$, which is in relatively good agreement with
model predictions for the contribution of disrupted GCs to the
stellar mass contained within the inner few kpc of the Galaxy
\citep[10$^7-$10$^8$~M$_\odot$, see, e.g.,][]{tr75,gn14}.  \cite{bk15}
showed that a similar mass in dissolved GCs can explain the gamma
ray detections by the {\it Fermi} satellite within 10$^{\circ}$ of
the Galactic centre.


An alternative mass estimate can be obtained from consideration
of a detailed model for the inner Galaxy, such as the one by
\cite{ro14}, which matches stellar counts in the 2MASS and SDSS
catalogs.  According to \cite{ro14} a cylinder of 2~kpc radius and
4~kpc height centred on the Galactic centre, contains a total $1.1
\times 10^{10}~M_\odot$.  There are a total of 3,244 APOGEE stars
within the same volume, 45 of which belong to the N-rich population.
Folding in factors accounting for the FG/SG ratio and maximum
mass-transfer binary contribution, we conclude that the total mass
of disrupted GCs would be $1.7 \times 10^8~M_\odot$, which is within
the range of estimates provided above.  

\subsubsection{Dissolved vs Existing GCs} \label{minexist}

Significantly, the mass contained in dissolved GCs is a factor of
$\sim$ 6-8 higher than the total mass in all {\it existing} Galactic
GCs \citep[$\sim 2.8 \times 10^7 M_\odot$,][]{kp09}.  This is
obviously an important result.  On the theory that N-rich stars are
byproducts of GC destruction, we would conclude that the Galactic
GCs are remnants of a formerly much larger GC system---or of what
would have become a much larger GC system today---that was largely
destroyed through interaction with the environment.  

This result naturally prompts us to ask whether such a dramatic
destruction rate is a common phenomenon in the universe, or whether
the Milky Way is in some way special.  \cite{hu14} reported that
the ratio $\eta$ between the mass of the GC system and total galaxy
mass (including both dark and stellar matter) is $\sim 4 \times
10^{-5}$, and showed that it is constant over several orders of
magnitude in galaxy mass and with a relatively small intrinsic
scatter of only 0.2 dex.  If the Galaxy underwent abnormally intense
GC destruction, that should manifest itself by a substantial
displacement from this mean value.  Adopting the Galactic GC system
mass from \cite{kp09} ($\sim 2.8 \times 10^7 M_\odot$) and a total
mass of the Galaxy ranging between $6 \times 10^{11} M_\odot$ and
$3 \times 10^{12} M_\odot$ \citep[e.g.,][]{bar14,fa15}, we obtain
$\eta_{MW}$ ranging between $9 \times 10^{-6}$ and $5 \times 10^{-5}$.
Considering the uncertainties in the numbers involved, one would
conclude that $\eta_{MW}$ is rather typical (if perhaps a bit on
the low side), which suggests that GC destruction at the levels
inferred from our results is a universal process.  The fact that
such a high rate of GC destruction is so finely tuned over a large
range of Galaxy masses and types is quite remarkable, and should
be the subject of careful theoretical examination.


\subsubsection{Bulge, thick disk, or halo?}
\label{halo}



As pointed out in Section~\ref{intro}, all components of the
Galaxy contribute to the stellar mass within its inner few
kpc.  Thus, what we call our ``bulge'' sample is in fact the
superposition of all stellar populations lying within the range of
Galactic coordinates and distances specified in Section~\ref{data}---which
certainly includes halo, thin and thick disk, bar and perhaps a
classical bulge.



The MDFs in Figure~\ref{mdfs} provide clues as to the nature
of the N-rich stars in our sample.  As pointed out in Section~\ref{lowmet},
the MDF of the N-rich population and that of the rest of the sample are
very different.  The bulge MDF extends towards above solar metallicity,
whereas that of the N-rich population peaks at [Fe/H]$\sim$--1,
suggesting in fact an association with the thick disk or halo
\citep[e.g.,][]{ne13a,ro14}.  Examination of the incidence of N-rich
stars in other regions of the Galaxy can help decide between these
two possibilities.  An association with the Galactic halo or
thick disk can be tested by searching for N-rich stars in other
parts of these Galactic components, and checking whether the observed
numbers match expectations based on the frequency of N-rich stars
in the inner Galaxy.  We start by comparing our numbers with those
obtained by other groups from analysis of SDSS-SEGUE data for halo
stars at larger Galactocentric distances.  \cite{ma11} and \cite{sc11}
estimated the total contribution of GC stars to the halo mass budget
(10-20\%) that is very similar to that resulting from the minimal
scenario.  However, their estimate is based on a much larger
primordial FG/SG ratio, which is far more compatible with
that assumed for our maximal scenario (see below).  Assuming
there is no strong reason for one to adopt different FG/SG
ratios for inner and outer halo, one would end up with a substantial
variation in the contribution of the halo stellar mass by GC stars
as a function of Galactocentric distance.  Adopting the FG/SG
ratio from the minimal scenario, the contribution to the halo mass
inferred from \cite{ma11} and \cite{sc11} would be reduced by a
factor of a few to several.  Indeed, in a more recent effort
based on APOGEE DR12 data, \cite{ma16} searched for N-rich stars
in high-latitude halo fields adopting a definition that is consistent
with that described in Section~\ref{data}.  From a resulting sample
of 5 halo N-rich stars, they concluded that, adopting a FG/SG ratio
consistent with our minimal scenario, the contribution of dissolved
GCs to the halo mass would be $\sim$4\%.  In conclusion, there are
fewer N-rich stars in the DR12 APOGEE halo sample, by a factor of
$\sim$ 5, than expected if the frequency of those stars was the
same across the entire halo.

The above approach suffers from a basic limitation stemming
from the small relative size of the APOGEE halo sample.  We exploit
the much larger APOGEE sample at low Galactic latitudes for a
statistically more robust comparison between expected and observed
N-rich star numbers.  For that purpose, we perform the following
exercise.  We first use models to estimate the ``component-specific''
frequency of N-rich stars under the assumption of their association
to each of those components, then use that frequency to estimate
the expected number of N-rich stars in other regions of the Galaxy.
By ``component-specific'' frequency, we mean the fraction of the
halo or thick-disk stars that are N-rich if the N-rich stars found
in the inner Galaxy are assumed to be associated entirely with with
either of those components.  This estimate requires knowledge of
the breakdown of the stellar mass in the inner Galaxy among various
Galactic components. For that purpose, we adopt the Besan\c con
models by \cite{ro12,ro14}, which match stellar counts in 2MASS and
SDSS by considering a combination of four components: a thick disk,
a thin disk, a halo, and a bar.  No classical bulge was needed in
order to fit the data for the inner Galaxy.  The mass breakdown
among the various components within a cylinder with 2~kpc radius
and 4~kpc height located at the Galactic centre is as follows:

\begin{itemize}

\item Thick disk: $5.7 \times 10^9~M_\odot$

\item Bar: $4.3 \times 10^9~M_\odot$

\item Halo: $8.5 \times 10^8~M_\odot$

\item Thin disk: $1.1 \times 10^8~M_\odot$

\end{itemize}

Next, using survey simulations based on the same models we calculate
how many N-rich stars are expected in the APOGEE sample of low
latitude fields with $|b|<20^\circ$ and $20^\circ < l < 340^\circ$ if
they were associated with either thick disk or halo.  The expected
number of N-rich stars integrated within all that area of sky is
given by:

\begin{equation}
N_{\rm exp} \,=\, N_{\rm obs} \, F(H|TD) \, f_{\rm Nr}(H|TD)  \label{eq1}
\end{equation}

\noindent where $N_{\rm obs}$ is the total number of stars observed
by APOGEE within that area, $F(H|TD)$ is the fraction of those stars
belonging to either the thick disk or the halo, and $f_{\rm Nr}(H|TD)$
is the component-specific frequency of N-rich stars in that area
of the sky.  We take $F(H|TD)$ from survey simulations based on the
Besan\c con models and estimate $f_{\rm Nr}(H|TD)$ from a combination
of the observed frequency of N-rich stars in our bulge fields and
model estimates for the contribution of thick-disk and halo stars
to the volume sampled, so that

\begin{equation}
f_{\rm Nr}(H|TD) \,=\, \frac{N_{\rm bulge,Nr}}{N_{\rm bulge}\, f_{\rm bulge}(H|TD)}
\label{eq2}
\end{equation}

\noindent where $N_{\rm bulge}$ is the number of stars observed
within the APOGEE bulge fields, $N_{\rm bulge,Nr}$ is the number
of those among the latter who are N-rich, and $f_{\rm bulge}(H|TD)$
is the component-specific frequency of N-rich stars in the inner
Galaxy, according to survey simulations.  For this exercise, we
limit the sample to stars with [Fe/H]$ < -0.7$, where the contribution
by halo and thick disk populations is maximal.

In Table~\ref{expectednumbers} numbers for the quantities defined
in equations~\ref{eq1} and \ref{eq2} are provided.  The total number
of stars in the APOGEE sample within $20 < l < 340^\circ$i and $|b|
< 20$ with stellar parameters as defined in Section~\ref{data}\footnote{The
simulations were actually carried out assuming $\log g  = 2$, but
that has no impact on our results, given APOGEE's relatively bright
(H=11.2) magnitude limit in bulge fields.}, and [Fe/H]$ < -0.7$ is
$N_{obs} = 476$.  The size of the APOGEE bulge sample in the same
metallicity range is $N_{bulge} = 742$, and the number of N-rich
stars is $N_{bulge,Nr} = 42$.

\begin{table} 
 \centering
  \caption{Expected numbers of N-rich stars with [Fe/H]$ < -0.7$ in other fields.}
  \begin{tabular}{ccccc}
  \hline
   &   $f$   &  $f_{Nr}$  & $f_{bulge}$ & $N_{exp}$  \\
 \hline
 Halo  &  0.021  &  0.07 $\pm$ 0.01  & 0.18  & 6 $\pm$ 1.5 \\
Thick Disk  &  0.920  &  0.38 $\pm$ 0.08  & 0.85  &   29 $\pm$ 7 \\
\hline
\end{tabular}
\label{expectednumbers}
\end{table}

The numbers in Table~\ref{expectednumbers} inform us that, if our
N-rich stars were associated with the thick disk, roughly 30 N-rich
stars should have been detected in other Galactic longitudes.  The
number in the case of an association with the halo is much smaller,
on order of 6.  As mentioned elsewhere, no N-rich stars were
found anywhere in the Galactic regions considered in this exercise.
These numbers tempt the assertion that a halo association is more
likely, as the total number of detections predicted in that case
is substantially smaller, and thus closer to the observed number (zero),
however both estimates are off from the observed values by 4~$\sigma$.
Such a substantial discrepancy may be simply due to our assumption
that $f_{Nr}$ inferred in the inner Galaxy applies everywhere else.
Previous work has indeed suggested the presence of a possible
gradient in the incidence of N-rich stars \citep{car13} in the
Galactic halo.  Considering also the fact that stars with second
generation abundances have been found in the halo
\citep[e.g.,][]{ma11,ca10,car13,li15,ft16}, it seems natural to conclude
that an association of our N-rich population to the inner halo seems
more likely than to the thick disk.  The discrepancy with the numbers
expected in other regions of the halo \citep[e.g.,][]{ma11,ma16}
indicates that the frequency of N-rich stars may be higher in the
inner halo, which is in and of itself an important constraint on
models for the origin of this population.

Finally, a by-product of the above exercise is an assessment of the
contribution of dissolved GCs to the stellar mass budget of those
Galactic components in the case where the N-rich stellar population
is associated to each of them.  We have seen in Section~\ref{minimal}
that, according to the minimal scenario and assuming maximum
contribution to the N-rich sample by binary stars, dissolved GCs
contribute $1.6-2.1 \times 10^8~M_\odot$ to the volume sampled
by the above model.  Assuming an association to the Galactic halo,
we would conclude that dissolved clusters contribute $\sim$~19-25\%
to the stellar mass of the halo within about 2~kpc of the Galactic
centre.  Conversely, if N-rich stars are members of the thick disk,
their contribution to the total mass of that Galactic component
would range between 2.8 and 3.7\%.

\subsection{The Maximal Scenario}  \label{maximal}


We now consider the {\it maximal scenario}, according to which GCs
were much more massive in the past and the vast majority of
the mass lost was in the form of FG stars.  In that scenario, SG
stars, such as the N-rich population reported in this paper, are
but a trace of the total GC population.  According to some models,
to satisfy observations of stellar abundances in Galactic GC members,
GCs would have to have been 10-100 times more massive in the past
\citep[for references, see, e.g.,][]{gr12,bl15,cz15}, which would
lead us to conclude that the bulk, if not all, of the stellar mass
of the bulge or inner Galaxy resulted from the dissolution
of GCs similar to the ones that populate the Galactic halo, bulge
and thick disk today, or their parent systems.  According to that
scenario \citep{sc11,gr12}, the presence of SG stars in the field
of the halo \citep{mg10} would similarly imply that a substantial
fraction of the halo was also built from dissolution of GCs
\citep{car13}.

To begin with, the MDFs of N-rich stars and the rest of the
inner Galaxy population are difficult to reconcile with the premise
that dissolved GCs contribute importantly to the {\it total} stellar mass
in the spatial region sampled in this work.  One possible way to
salvage that proposition is by assuming that the FG/SG in GCs
increases substantially towards high metallicity.  Statistics on
C-N anticorrelations in metal-rich ([Fe/H] $\simgreater$--0.5) GCs
are currently meager, which makes it difficult to test the hypothesis.
However, \cite{ca10} reported that there is tentative evidence for
the presence of a correlation between GC metallicity and the extension
of the Na-O anticorrelation and, by association, the incidence of
N-rich-like, SG stars in Galactic GCs.  Moreover, studying
M31 GCs in integrated light, \cite{s13} established a correlation
between mean [N/Fe] and [Fe/H], again arguing for an enhancement,
rather than a diminution, in the relative number of N-rich stars
in more metal-rich GCs.  Therefore, if GCs had contributed
substantially to the mass of the bulge, we would expect the N-rich
population to have a much more metal-rich MDF.  Therefore, we
conclude that, at face value, the MDFs suggest that it is unlikely
that dissolved GCs contribute significantly to the total stellar
mass within the volume sampled in this study.

An important contribution of dissolved GCs to the stellar mass
budget is also likely incompatible with current understanding of
the structure and stellar content of the inner Galaxy.  Regarding
stellar population content, there may be small, but non-negligible,
differences in abundance pattern between first-generation stars in
{\it existing} GCs and both halo and bulge field populations.  For
instance, Figure 3 of \cite{ca09} hints at the possible existence
of small [O/Fe] differences between field and GC populations in the
halo, and larger ones in the bulge.  Whether these are real or due
to systematics stemming from differences in data quality and/or
analysis methods is not clear.  As regards the structure of the
bulge, a substantial part of its stellar mass is seemingly contained
in the bar \citep[e.g.,][]{ra14,ne13a}.  Indeed, based on the Besan\c
con models \citep{ro14}, one would estimate that the bar contributes
$\sim$ 40\% of all stellar mass within 2 kpc of the Galactic centre,
which implies that a considerable part of the bulge population
probably resulted from secular evolution of the disk.


The above caveats aside, in this Section we define the maximal
scenario in such a way that the constraints from low-metallicity
stars discussed in Section~\ref{lowmet} are met.  The resulting
numbers are then used to compare the total mass in dissolved GCs
with that of the existing Galactic GCs, and also to estimate the
maximal contribution of dissolved GCs to the stellar mass budget
of various Galactic components.  The key constraint posed by the
MDF of N-rich stars of [Fe/H] $\simless$ --1 is that the FG/SG ratio
in dissolved GCs cannot exceed 0.93.  Thus, if the ratio between
N-normal and N-rich stars in dissolved GCs is maximum, then the 45
N-rich stars contained in the cylindrical volume considered in
Section~\ref{minimal} correspond to only 7\% of the mass in that
volume.  Therefore, the total number of stars resulting from GC
destruction in our sample should be 843.  Considering that our
sample contains 5,140 stars, the maximal scenario implies that
$\sim$~16\% of all stars in our bulge sample result from dissolution
of Galactic GCs.  In this scenario, the mass in dissolved GCs
outnumbers that of the existing systems by a factor of 
$\sim$~60--90, depending on the total mass adopted for the inner
Galaxy (Section~\ref{minimal}).  If the numbers used considered
only stars within the cylindrical volume included in the Besan\c
con model calculations, $\sim$~20\% of the stars in that cylinder
would have resulted from GC dissolution, leading to a total mass
that outweighs the GC system by a factor of 80.

To estimate the contribution of dissolved GCs to the stellar
mass in the halo or thick disk, we simply compare the latter
percentage with those based on the Besan\c con models, listed in
Section~\ref{halo}.  According to those models, the Galactic halo
accounts for $\sim$~8\% of the stellar mass in the inner Galaxy,
so that the maximal mass in dissolved GCs outweighs that in the
inner halo by over a factor of 2.  Conversely, considering an
association of N-rich stars to the thick disk, which according to
the Besan\c con models accounts for $\sim$~50\% of the mass in the
inner Galaxy, one would be led to conclude that the $\sim$~40\% of
the mass in the thick disk results from GC dissolution.  According
to the maximal scenario, the halo alone cannot contribute to the
whole mass of dissolved GCs, with the thick disk contributing an important
fraction, if not all of it.




In conclusion, the evidence accumulated thus far seems to suggest
that the maximal scenario is ruled out by the data, and the 
FG/SG ratio in GCs was much lower than the 93\% limit described
above.  The latter implies that the contribution of dissolved GCs
to the halo mass inferred in previous studies is overestimated.  It
seems safe to conclude that the fraction of halo stellar mass
contributed by dissolved GCs peaks towards the inner halo, but the
exact number is hard to pin down due to uncertainties in the 
FG/SG ratio, in the contribution of the inner halo to the stellar
mass within the inner few kpc of the Galaxy, and in the association
of the N-rich stars to any of the overlapping components of the
Galaxy in its inner regions.

\subsection{MDFs of N-rich stars vs Galactic GCs} \label{nvsgcs}

We now examine the comparison between the MDFs of the N-rich stars
and the Galactic GCs, shown in the bottom panel of Figure~\ref{mdfs}.
As is well known, the Galactic GC MDF is seemingly bimodal, showing
evidence for the presence of two peaks, at [Fe/H] $\sim$ --1.6 and
--0.7, with a trough at [Fe/H]$\sim$--1, which is precisely where
the peak of the N-rich MDF is located.  Low number statistics
prevents an analysis of the shape of the N-rich star MDF.  For
instance, as noted above, the apparent trough in the MDF of
the N-rich population at [Fe/H]$ \sim -0.9$ has no statistical
significance.  Nevertheless, despite the relatively low numbers,
much can be learned from comparison between the N-rich star MDF and
that of the GC system.  The range encompassed by the N-rich MDF
goes from [Fe/H]$\sim$--1.5 to almost solar.  In comparison, the
GC MDF extends towards metallicities as low as [Fe/H]$\sim$--2.5.
It is unclear whether the lack of stars with [Fe/H] $\simless$ --1.5
in the N-rich MDF is real or due to low number statistics.
As discussed in Section~\ref{bulgemdf}, the APOGEE sample
studied in this paper presents no important bias against metal poor
stars.

The Galactic GC MDF differs substantially from that of the N-rich
population even when only GCs within the volume containing the
N-rich sample (Section~\ref{data}) are considered, as indicated by
the hatched gray histogram on the bottom panel of Figure~\ref{mdfs}).
Despite the relatively low numbers, the bulge GC MDF is markedly
different from that of the N-rich population, with no peak at any
particular metallicity.  A Kolmogorov-Smirnoff test strongly rejects
the hypothesis that the two samples are drawn from the same parent
population, at a level $P = 0.002$.


It is clear from the above that the N-rich star MDF does not match
that of the existing GC system, either considered as a whole, or
taking the halo and bulge/thick disk components separately.  We
interpret this result as evidence that evaporation of existing
Galactic GCs has not contributed significantly to the population
of N-rich stars.  The rationale behind this interpretation is that
it would be difficult, without tidal evaporation that is strongly
dependent on metallicity, to deplete the masses of an original
population of GCs by an order of magnitude, while completely
obliterating its MDF.  Tidal evaporation is of course likely to
have operated differently on GCs of different metallicity, because
more metal-rich GCs, typically located at smaller Galactocentric
distances, were probably subject to more vigorous tidal evaporation.
In that case, we would possibly expect the N-rich MDF to resemble
more strongly that of thick disk/bulge GCs, which we show not to
be the case.  In short, the MDFs require destruction to have been
very efficient for GCs in a narrow metallicity range around
[Fe/H]$\sim$--1, and less efficient everywhere else, which seems
contrived.

One additional possibility is that the MDF of the Galactic GC system
is not itself bimodal, but rather suffers from non-linearity effects
such as those claimed by \cite{yo06} to affect the conversion between
integrated colours and metallicity of extragalactic globular cluster
systems.  While giving it a nod in this paper, we choose to defer
this line of reasoning to a future study, and take the Galactic GC
MDF from Figure~\ref{mdfs} at face value.  We therefore suggest
that the N-rich population, if at all associated with GCs, was
predominantly produced by the {\it destruction} of a large population
of early Galactic GCs.  In fact, if indeed the halo and bulge/disk
components of the Galactic GC system had different origins
\citep[e.g.,][]{sh10,to13}, it is also likely that their
destruction efficiencies were different, so that one would indeed
{\it expect} that the MDF of the destroyed and surviving populations
differ.  In Section~\ref{origin} we discuss possible scenarios for the
origin and fate of this population of globular clusters, the remnants
of which we seem to have discovered in the inner Galaxy.


\subsection{Dissolved GCs in the Cores of Other Massive Galaxies?}
\label{ETGs}

While GCs and their parent populations do not seem to be important
building blocks of the Galactic bulge, the situation may be different
in other environments.  Unlike the inner Galaxy, where the
bar and thick disk contribute $\sim$ 91\% of all stellar mass
(Section~\ref{halo}) the cores of early-type galaxies are dominated
by a spheroidal component, which like the Galactic halo is the result
of accretion of a large number of low mass stellar systems.  We
have seen from the discussion above that the evidence points to the
N-rich population being associated with the Galactic halo, in which
case dissolved GCs contribute a {\it minimum} 19-25\% of the stellar
halo mass.  If dissolved GCs contribute a similar fraction to the
stellar mass in the cores of early-type galaxies, one would expect
that the mean abundances inferred from integrated-light studies to
be influenced by the chemical compositions typical of N-rich stars.
In that case, the abundances of elements such as nitrogen, sodium,
and aluminium could be enhanced in the integrated spectra of
early-type galaxies.

Interestingly, integrated light studies have shown that early-type
galaxies are characterised by large mean values of [N/Fe] and
[Na/Fe], which are moreover strongly correlated with galaxy mass
and velocity dispersion \citep{s07,gr07,co14,wo14,sm15}.  Moreover,
in a recent study \cite{zi15} estimated [Na/Fe]$\simgreater +0.3$
within the inner $\sim$~0.4 kpc of the Andromeda galaxy, with
[Na/Fe] possibly increasing to as much as +1.0 within the inner
$\sim$~40~pc.  Although difficult to disentangle from IMF effects
on the NaIi $\lambda 8200{\rm\AA}$ line \citep[e.g.,][]{st71,s00,vdc10},
this is a strong indication of the presence of Na-enhanced populations
in the core of Andromeda.  Along the same lines, \cite{bu84} showed
\citep[zero-point uncertainties aside, see][]{s12} that CN bands
in the core of M31 are enhanced, being consistent with their strengths
in the integrated spectra of M31 and Galactic GCs.  The work by
\cite{st13} is also worthy of notice in this context.  They analysed
the optical spectrum of an ultra compact dwarf galaxy satellite of
M~60, in the Virgo cluster, and found very high [N/Fe] and [Na/Fe]
abundance ratios.

We speculate that these results indicate the presence of a population
of dissolved GCs in the core of Andromeda, and possibly also in the
central regions of early-type galaxies.  The population of
N-rich stars we discovered in the heart of the Galaxy may thus be
the tracers of a global phenomenon associated with the formation
of spheroidal systems in general.


\subsection{Origin of the presumptive population of dissolved globular
clusters}  \label{origin}

The question of the origin of N-rich stars is inevitably tangled
with two major unsolved problems in galaxy formation, namely,
formation of the Galactic bulge (and galaxy bulges in general) and
globular cluster formation.  In this sub-section we briefly discuss
four possible scenarios to explain the origin of N-rich
stars.  The first two scenarios assume, as usual, that N-rich stars
were initially formed within, and later lost to, the gravitational
potential of globular clusters.  A third scenario relaxes this
assumption and contends that instead N-rich stars were formed within
the same molecular clouds as GCs, but were never gravitationally
bound to them.  A fourth scenario suggests that N-rich stars were
instead never necessarily associated with GCs and are rather the
oldest existing stellar population formed in the Galactic bulge
itself.

\subsubsection{GC Origin} \label{gcorigin}

It is now generally agreed that most globular clusters formed in
giant molecular clouds (GMCs) generated by disk instabilities in
galaxies at $z \simgreater 2$
\citep[e.g.,][]{kg05,sh10,to13,kr14,rh15,kr15}.  Globular cluster
formation motivated by mergers of gas-rich galaxies is another
possible mechanism \citep{az92,mu10}, but one thought to make a
minor contribution to the total stellar mass allocated in GC systems
today \citep{kg05}.

\cite{eh10} championed the notion that newborn clusters are efficiently
destroyed by tidal interaction with GMCs, which are present in the
very environment that initially gave birth to clusters---the so
called ``cruel cradle effect'' \citep[see also][]{kr11,kr12}.
\cite{kr14,kr15} proposed a model, based on an analytical formulation,
where GC formation takes place in two phases.  In the first phase,
GCs are formed with a power-law mass distribution, within GMCs
hosted by turbulent disks at $z \sim 2-3$.  Formation is followed
quickly by vigorous disruption due to tidal interaction with GMCs.
Since tidal destruction of lower-mass clusters is more efficient,
the original power-law mass function is converted into the lognormal
distribution observed today.  Survival of GCs is ultimately dependent
on the occurrence of a galaxy merger, which removes them from their
inhospitable birthplaces.  Mergers trigger a second phase, during
which GCs are incorporated into the halo of a new host galaxy or
merger remnant, where they suffer a more gentle, longer tidal
evaporation through interaction with the gravitational potential
of the new host galaxy.  This scenario matches a number of properties
of GC systems, including the GC specific frequency and mass
distribution in galaxies with a range of metallicities and halo/stellar
masses.  It is also in line with previous suggestions that the
disk/bulge GCs were formed {\it in situ} \citep{sh10}, whereas the
halo GC system has largely been accreted \citep{to13}.

This scenario for GC formation suggests that there may be at least
two possible channels for the production of N-rich stars involving
GC destruction:  {\it (i)} the {\it in situ GC} channel, whereby
these stars originate from the population of GCs that migrated into
the inner Galaxy from an early turbulent Galactic disk, being
destroyed in the process; {\it (ii)} the {\it ex situ GC} channel,
according to which these stars originate from the dissolution of a
population of accreted GCs.  We briefly discuss these formation
channels below.


\begin{enumerate}

\item {\it In situ GC origin:}  \cite{bo07} and \cite{el08} proposed
that the Galactic bulge was formed by the coallescence of giant
clumps hosted by turbulent disks at high redshift \citep[see review
by][]{bo15}.  Those clumps were also the sites of GC formation, as
proposed by various authors cited above.  According to this scenario,
one would expect that GCs initially formed in the disk and eventually
migrated towards the inner Galaxy, with some of them losing mass and/or
being destroyed in the process, and others surviving in the Galactic
thick disk and bulge.  In that scenario, the GCs associated with
the thick disk and bulge of the Galaxy today would be remnants of
an active past of star formation in the Galactic disk \citep{sh10}.
Presumably, this process would naturally result in the presence of
the byproducts of GC dissolution in the inner Galaxy,
as reported in this paper.  Moreover, one would also expect such
populations to be found in the Galactic disk, as a considerable
amount of tidal destruction is expected to have taken place during
interaction between GCs and GMCs in the disk \citep{kr15}.  

As reported in Section~\ref{halo}, a  search for N-rich disk
stars in the APOGEE DR12 database, adopting the same parameters
as described in Section~\ref{data}, but focusing instead on $20^{\circ}
< l < 340^{\circ}$, resulted in no field stars with an N-rich
abundance pattern.  This result does not necessarily mean that
N-rich stars do not exist in the disk, since, despite its very large
sample, the APOGEE coverage of the disk is of course limited.
Moreover, one would in any case expect relative numbers in the disk
to be lower than in the inner Galaxy, given that the disk has been subject
to a much longer history of star formation at later times, for
several Gyr, where conditions did not favor formation of globular
clusters, or their parent systems \citep{kr15}, leading to a decrease
of the ratio of N-rich/N-normal stars.

As discussed in Section~\ref{metdis}, the MDFs of the N-rich stars
and thick disk/bulge GCs are quite different, with the former peaking
at [Fe/H]~$\sim-1$ and the latter spanning a wide range of metallicities
and not peaking at any particular value.  \cite{sh10} pointed out
that the metallicities of star forming clumps in $z \sim 2-3$ disks
were high and thus not compatible with the formation of the metal-poor
component of the Galactic GC system.  It is unclear whether current
models for formation of GCs in turbulent disks can account for the
existence of the metal-poor GCs and N-rich stars seen in the inner
Galaxy.  \cite{sh10} suggest that metal-poor GCs may have formed
along cold filaments that were proposed by \cite{de09} and others
to account for high star formation rates in the early universe.
Whether the MDF of the existing GCs and that of those that were
dissolved in the past can be accounted for in detail by these models
is an open question.  Another possibility for the formation of these
metal-poor GCs is discussed next.



\item {\it Ex situ GC origin:} Following \cite{kr15}, globular
clusters can also have been formed in lower mass galaxies, in much
the same way as described above, and later accreted with their host
galaxy into the deeper potential well of the Galactic dark matter
halo, where they could have been subject to tidal evaporation or
destruction.  There is abundant evidence for the formation of the
Galactic halo itself, and that of the nearest giant spiral galaxy,
Andromeda, through accretion of satellite galaxies
\citep[e.g.,][]{sz78,ib94,be06b,lm10,ib14,gi14}.  Moreover the tidal
evaporation of halo GCs through interaction with the Galactic
potential has been spotted in real time \citep[e.g.,][]{ro02,od03,be06a},
and must be responsible for at least part of the population of
N-rich stars detected in the halo by \cite{mg10} and \cite{ma11},
and possibly also for a small fraction of those found in the inner Galaxy.
However, tidal evaporation of this nature is probably too inefficient
\citep{bm03,kr15} to account for a substantial fraction of the
N-rich stars detected in the inner Galaxy.  Indeed, it is thought that GCs
accreted in this process make up the blue/metal-poor component of
GC bimodal distributions in galaxies \citep{to13}, so that their
characteristic metallicities are typically lower than those of the
N-rich stars identified in the inner Galaxy in this study (c.f.
Figure~\ref{mdfs}).  However, if one accepts the scenario whereby
GCs are formed in clumpy disks at $z \simgreater 2$, then it is
possible that low-mass galaxies accreted to the early Galaxy hosted
a population of GCs that was previously dissolved/stripped in their
parent turbulent disks, and were later incorporated into the field
of the Milky Way itself.  Dynamical friction could then drive the
stars---including those of the N-rich variety--- belonging to the
most massive of those systems, which potentially host the most
metal-rich GCs, towards the central parts of the Galaxy, where they
reside today.

\end{enumerate}

\subsubsection{Shared nursery origin:} \label{sharednursery}

The present paper reports the
finding that the vast majority of the stars with a so-called
SG star abundance pattern appears to reside in the
field, not in clusters (Section~\ref{dissolve}).  This result quite
naturally invites one to question the standard assumption that these
stars were formed in GCs to begin with.  Indeed the fact that they
were first discovered in GCs is due to a severe observational bias
and does not imply a {\it sine qua non} genetic link---at least not
until detailed physical models of GC formation account for their
existence in a quantitative sense, by matching the extant data sets.
The stark disagreement between the MDF of the N-rich
stars and that of the Galactic GC system---modulo possible zero-point 
differences---lends further support to the notion that the
two populations do not share the same origin.  

In light of this evidence, a model that can form N-rich stars outside
the gravitational well of GCs may be required.  The natural sites
for the formation of such field N-rich stars would be the very
molecular clouds that formed the GCs.  It is conceivable that not
all stars formed in those clouds ended up in GCs---indeed, the
evidence from studies of molecular clouds in the Galaxy point in
the opposite direction \citep[e.g.,][]{lo14}.  Stars not bound to
GCs would be lost to the field quite easily, thus making up the
majority of the field population observed today.  One could 
imagine a scenario where the ratio between GC-bound and unbound
stars formed in such molecular clouds is a function of the physical
conditions in the cloud (e.g., density, chemical composition) as
well as the environment.  Such a scenario could potentially explain
the predominance of N-rich stars that are currently not gravitationally
bound to any Galactic GC, without the need to invoke efficient GC
destruction.  Open clusters, having been formed under physical
conditions that are rather different from those that gestated GCs,
were not able to produce stars with a SG abundance
pattern.  If molecular clouds existed that formed N-rich stars, but
no GCs, this scenario would naturally account for the mismatch
between the MDFs of the N-rich stars and the existing Galactic GCs.
Moreover, if this strawman scenario is correct, one would expect
to find N-rich stars in the same environments as GCs today, seeing as
at least some of them would have been formed in the same molecular
clouds as the GCs themselves.  As discussed above, N-rich stars
were indeed discovered in the halo \citep{ma11,car13} and
in the inner Galaxy (this study), but it is noteworthy that none so far
has been identified in the thick disk.

One remaining issue that is not simply solved by this scenario is
the mass budget problem, as the FG/SG ratio in the stellar
populations below [Fe/H] $\simless$ --1 is strongly constrained by
the bulge MDF discussed in Section~\ref{metdis}.  This problem could
possibly be circumvented if molecular clouds formed stars in a range
of [Fe/H], which might make possible pollution of the material going
into the formation of metal poor N-rich stars by the ejecta of their
more metal-rich counterparts.  Admittedly, this seems a bit contrived,
but perhaps not entirely outside the realm of possibilities, since
evidence for self enrichment in star forming regions has been found
before \citep{cl92}.  Assuming there is any physical reality to
these speculations, one could conceivably devise differences between
the abundance patterns of field and GC N-rich stars that would hint
at their origin, and possibly help constrain the models for the
formation of both stellar populations.

\subsubsection{Oldest stars:}  \label{oldestars}

We conclude our exploration of scenarios that may explain the origin
of N-rich stars by briefly mentioning another possible interpretation
not associated with a globular cluster origin.  For a more detailed
discussion, we refer the readers to a forthcoming paper \citep[][in
preparation]{ch16}.  This scenario advocates that N-rich stars are
among the oldest in the Galaxy and their abundances are in fact the
imprints of the very early chemical enrichment by the first stellar
generations, which polluted the interstellar medium prior to the
formation of GCs.  Some numerical simulations predict that the
oldest stars in the Galaxy are indeed to be found in its central
regions \cite[e.g.,][]{br07,tu10}.  In the Galactic halo these very
early phases of chemical enrichment are traced by halo stars with
[Fe/H]~$\simless-2.5$.  In the central regions of the Galaxy however,
the star formation rate is believed to have been higher, so that
the oldest stars in the bulge would have [Fe/H]~$\sim-1$ \citep{ch11}
which is around the metal poor tail of the old bulge MDF, and also
where the N-rich MDF peaks (Section~\ref{metdis}).  According to
this view, N-rich stars may be opening a window into the initial
stages of the formation of the Galaxy, which is of course a very
exciting prospect.

There are similarities between the abundance patterns of N-rich
stars and model predictions for the oldest stars.  For instance,
models based on enrichment by fast rotating stars \citep[the so
called ``spinstars'', see][for a review]{ch13} predict an enhancement
in $^{14}$N and $^{13}$C, a correlation between [N/Fe] and [Al/Fe],
and a modest enhancement in $^{12}$C, as well as some contribution
to the $s$-process nucleosynthesis.  While the nitrogen enhancement
and its correlation with aluminium are corroborated by the data,
observations at face value are at odds with predictions for carbon,
which is depleted and in fact anti-correlated with nitrogen in the
N-rich sample.  For the N-rich stars from our sample, it is hard
to assess without detailed modeling of the impact of mixing, whether
the latter disagreement is real or due to stellar evolution effects.
On the other hand, recall that we removed stars with [C/Fe]~$>$~+0.15
from consideration, which obviously biased our sample against stars
with strong carbon enhancements.  Determination of s-process element
abundances for a sample of N-rich stars including some with high
[C/Fe] would provide a good test of those model predictions.


\subsubsection{Final Considerations} \label{final}

Without further information, it is impossible to decide which of
the above scenarios is the most likely to account for the discovery
reported in this paper.  The one that perhaps is the least favoured
by the data is that proposing that N-rich stars are the oldest in
the Galaxy, in view of the prediction of enhanced carbon abundances,
which is not verified by the data.  On the other hand, this
scenario may not suffer from the mass budget problem.  It is however
conceivable that all these channels have contributed to the formation
of the N-rich stellar population present in the inner Galaxy.

The {\it in situ} and {\it ex situ} GC channels described above
assume that all N-rich stars were necessarily formed in Galactic
GCs and were later lost to the field when the GCs were ultimately
disrupted.  Each channel likens the newly found stars to existing
Galactic GCs associated to the disk/bulge ({\it in situ}) and halo
({\it ex situ}) components of the Galaxy.  In the {\it shared
nursery} scenario, there should be a similar balance between an
{\it in situ} and an {\it ex situ} origin for the N-rich stars
currently found inhabiting the inner Galaxy.  That is because in this
scenario N-rich stars are proposed to have been gestated in similar
molecular clouds as GCs, thus sharing their environmental origin.
On the other hand, the ``oldest stars'' channel requires
these stars to have been formed in situ, as only in the deep graviational
potential of the Galaxy could star formation be intense enough to generate
stars with [Fe/H]~$\sim$~--1 in a very short time.

We hypothesise that some yet unknown fraction of the [Fe/H] $<$ --1
N-rich stars found in the inner Galaxy originated {\it ex situ}, whereas
the metal-rich component and the remainder of the metal-poor stars
would have been formed {\it in situ} with perhaps some contribution
by star formation at very early stages.  The term {\it in situ}
here means different things for these two latter populations, as
dissolved GCs are proposed to have been formed in an early turbulent
disk and later migrated to the inner Galaxy \citep{sh10}, whereas the
``oldest stars'' are proposed to having been formed in the bulge.
A more detailed analysis of an enlarged sample, including further
elemental abundances, will surely provide further insights into how
the parent systems were formed, potentially destroyed, and eventually
left these trace populations as remnants in the inner Galaxy, and
how much stellar mass was contributed by the mechanisms put forward
here---or by other unforeseen means.  By tracing back the steps
that brought these populations to their current configuration, we
expect to gain a deeper understanding of the processes leading to
the formation of globular clusters, and of the Galactic bulge itself.


\section{Summary}  \label{summary}

The main conclusions of this paper are the following:

\begin{itemize}

\item We have discovered a number of field stars in the inner Galaxy
with high [N/Fe], which in addition is anti-correlated
with the abundance of carbon and correlated with that of aluminium.
This abundance pattern is characteristic so called ``second
generation'' globular cluster stellar populations.


\item The lower limit of the stellar mass ascribed to this new
stellar population with a GC-like abundance pattern exceeds that
of the existing GCs by a factor of $\sim$ 6--8.  If these stars are
assumed to be the by-products of the destruction of old globular
clusters, our result implies that the Galactic GCs are the remnants
of a much larger system that was largely destroyed.  If the GC
origin of the N-rich stars is confirmed, the location of the Galaxy
very near the mean ratio $\eta$ between integrated GC mass and total
mass from \cite{hu14} suggests that vigorous GC destruction is a
universal process.  That notion is further corroborated,
although tentatively, by similarities between the mean abundance patterns
in the cores of early-type galaxies and those of N-rich stars.  It
is striking that such a large destruction rate is so finely tuned
across a wide range of masses.

\item Again assuming that N-rich stars result from GC dissolution,
we derive an {\it upper limit} of $\sim$ 93\% for the fraction of
GC mass in the form of stars with a FG abundance pattern.
This result challenges models of chemical evolution of globular
clusters that postulate larger fractions.  It makes it quite
difficult, perhaps impossible, to solve the mass budget problem
without revising yields from stellar evolution models.  Failing
that, the whole notion that ``second generation'' stars are connected
to their ``first generation'' counterparts by chemical evolution
through incorporation of byproducts of stellar nucleosynthesis may
have to be altogether dropped.  That is is a riveting prospect.

\item Under reasonable assumptions for the primordial ratio between
first- and second-generation stars the contribution of dissolved
globular clusters to the mass of the Galactic bulge is estimated
to be small, of the order of a few percent.  Given the spatial
overlap of all Galactic components within the central few kpc of
the Galaxy, a definition regarding whether N-rich stars belong to the halo,
the thick disk, or other components is impossible.  The evidence
discussed in this paper favors an association to the inner halo of
the Galaxy, although this conclusion is by no means definitive.  If
N-rich stars indeed belong to the halo, they contribute a minimum
of 19-25\% to the stellar mass in its inner $\sim$~2~kpc.  An accurate
estimate is hampered by uncertainties in quantities such as the
FG/SG ratio, the contribution of the halo to the stellar mass
in the inner Galaxy, and the fraction of N-rich stars that are
associated with the inner halo.  Nevertheless, this mass fraction
exceeds that found in other studies for the outer halo by a factor
of several.

\item The metallicity distribution function of the newly discovered
stellar population does not match that of the bulk of the Galactic
bulge.  This result suggests that the bulge cannot have been built
from a population of dissolved globular clusters, unless special
assumptions are made regarding the incidence of N-rich populations
in metal-rich GCs, which are not in agreement with current observational
evidence.  We conclude that destruction and/or evaporation of
globular clusters accounts for no more than a few percent of the
stellar mass of the Galactic bulge.  

\item The metallicity distribution function of the newly discovered
stars does not match that of the Galactic GC populations.  This
result suggests that the N-rich stellar population discovered in
the inner Galaxy does not result from simple ``thinning'' of the
existing Galactic globular cluster population through mass loss
from tidal evaporation.  It may also imply that these stars have
never been gravitationally bound to any globular cluster.

\item We hypothesise that N-rich stars may have resulted from four
possible mechanisms.  The first two are closerly related and involve
the destruction of GC populations formed {\it in situ} and {\it ex
situ}.  A third mechanism contends that N-rich stars (and presumably
most field stars with a so-called second generation abundance
pattern) were not perforce associated with Galactic GCs, and perhaps
most of them may have been formed in similar environments, while
never being gravitationally bound to the GCs themselves.  A fourth
mechanism, namely, a very early star formation from a medium polluted
by ``spinstars'' matches some of the data at least qualitatively,
but more work needs to be done to put its predictions to test.

\item Regardless of their origin, we find that the vast majority
of stars with a second generation abundance pattern today live
in the field, not in globular clusters.  If indeed these newly
discovered stars were never associated with any globular cluster,
it is conceivable that the abundance patterns of these ``field
second generation stars'', may differ in detail from those of
their globular cluster counterparts, in ways that need to be
theoretically devised and observationally verified.

\item We emphasise that there are a number of key pieces of evidence
that are difficult to understand, including the absence, in our sample,
of N-rich stars in the Galactic disk, the MDF differences between
N-rich stars and the Galactic GC system, and whether the relative
numbers of N-rich stars in the inner and outer halo can be reconciled
within a single formation scenario.  The ultimate association of
the N-rich stars to their counterparts currently living in Galactic
GCs also requires the determination of other elemental abundances,
such as Na and $s$-process elements.  Detailed modelling and more
extended observations will hopefully address these questions.

\end{itemize}

The results presented herein provide eloquent confirmation of the
power of high-resolution spectroscopy applied to large stellar
samples to provide key insights into the history of formation of
the Galaxy.  In the H band, the added benefit of low extinction and
relatively easy access to lines of CN, CO, and OH make APOGEE a
powerful tool to identify remnants of globular cluster disruption
at low Galactic latitudes.  Exploration of methods to constrain the
contribution by various parent systems to the stellar field of the
Galaxy is a growing field, which we expect will flourish within the
next several years, with the delivery of larger stellar samples,
with even better and more detailed phase space and chemical
information.


\section*{Acknowledgments}

R.P.S. thanks Diederik Kruijssen, Alessio Mucciarelli, Carmela
Lardo, Maurizio Salaris, Rob Crain, Renyue Cen, Jenny Greene, David
Spergel, and Jakob Walcher for enlightening discussions and/or
comments on an early version of this manuscript, and Ingrid for
everything.

Funding for SDSS-III has been provided by the Alfred P. Sloan
Foundation, the Participating Institutions, the National Science
Foundation, and the U.S. Department of Energy Office of Science.
The SDSS-III web site is {\tt http://www.sdss3.org/}.  SDSS-III is
managed by the Astrophysical Research Consortium for the Participating
Institutions of the SDSS-III Collaboration including the University
of Arizona, the Brazilian Participation Group, Brookhaven National
Laboratory, University of Cambridge, Carnegie Mellon University,
University of Florida, the French Participation Group, the German
Participation Group, Harvard University, the Instituto de Astrof\'\i
sica de Canarias, the Michigan State/Notre Dame/JINA Participation
Group, Johns Hopkins University, Lawrence Berkeley National Laboratory,
Max Planck Institute for Astrophysics, New Mexico State University,
New York University, Ohio State University, Pennsylvania State
University, University of Portsmouth, Princeton University, the
Spanish Participation Group, University of Tokyo, University of
Utah, Vanderbilt University, University of Virginia, University of
Washington, and Yale University.
T.C.B. acknowledges partial support for this work from grants PHY
08-22648; Physics Frontier Center/Joint Institute or Nuclear
Astrophysics (JINA), and PHY 14-30152; Physics Frontier Center/JINA
Center for the Evolution of the Elements (JINA-CEE), awarded by the
US National Science Foundation.  R.C. acknowledges support provided
by the Spanish Ministry of Economy and Competiviness under grants
AYA2010−16717 and AYA2013−42781P.
C.A.P. is thankful for support from the Spanish Ministry of Economy and
Competitiveness (MINECO) through grant AYA2014-56359-P.
S.L.M. acknowledges the support of the Australian Research Council
through DECRA Fellowship DE140100598
Szabolcs M{\'e}sz{\'a}ros has been supported by the J{\'a}nos Bolyai
Research Scholarship of the Hungarian Academy of Sciences.

\clearpage

\appendix  

\section{N-rich stars}
In Table~\ref{list} we provide the identities of our sample of 58
stars identified as N-rich in the inner Galaxy by APOGEE, along
with parameters employed in the analysis.

\begin{table*} \label{list}
 \centering
  \caption{N-rich stars identified in the inner Galaxy.}
  \begin{adjustbox}{max width=\textwidth}
  \begin{tabular}{cllrrcccccr}
  \hline
APOGEE\_ID  & $\alpha_{2000}$ & $\delta_{2000}$ & S/N &  $T_{\rm eff}$  &
$\log g$ &  
{\it [Fe/H]} & 
{\it [Al/Fe]} & 
{\it [C/Fe]} & 
{\it [N/Fe]} & 
d (kpc) \\
 \hline
  2M16493657-2028146 & 252.402403 & -20.47073  & 152 & 4454  & 1.3  &
  -1.15 $\pm$ 0.05  &-0.06  $\pm$ 0.15  & -0.37  $\pm$ 0.10  & 0.78 $\pm$ 0.13
      &7.1  $\pm$ 1.6  \\
  2M16514646-2127071 & 252.943621 & -21.451977 & 121 & 4439  & 1.7  &
  -0.86 $\pm$ 0.04  & 0.36  $\pm$ 0.13  & -0.21  $\pm$ 0.08  & 0.10 $\pm$ 0.11
      &5.3  $\pm$ 1.3  \\
  2M17024730-2210387 & 255.697092 & -22.177443 & 211 & 4296  & 1.3  &
  -0.95 $\pm$ 0.04  & 0.36  $\pm$ 0.11  & -0.41  $\pm$ 0.08  & 1.01 $\pm$ 0.11
      &5.1  $\pm$ 1.1  \\
  2M17134700-2441353 & 258.445841 & -24.693153 & 133 & 4298  & 0.9  &
  -1.55 $\pm$ 0.05  & 0.35  $\pm$ 0.20  & -0.42  $\pm$ 0.12  & 0.82 $\pm$ 0.15
      &8.0  $\pm$ 1.3  \\
  2M17161691-2458586 & 259.070484 & -24.982969 & 158 & 4133  & 1.3  &
  -0.10 $\pm$ 0.03  & 0.34  $\pm$ 0.06  & -0.30  $\pm$ 0.04  & 0.59 $\pm$ 0.07
      &5.7  $\pm$ 1.6  \\
  2M17173203-2439094 & 259.383468 & -24.65262  & 223 & 3780  &-0.2  &
  -0.78 $\pm$ 0.04  & 0.28  $\pm$ 0.08  & -0.34  $\pm$ 0.05  & 0.58 $\pm$ 0.08
      &7.4  $\pm$ 0.5  \\
  2M17193271-2732214 & 259.886313 & -27.539303 & 219 & 3971  & 0.2  &
  -1.30 $\pm$ 0.04  &-0.30  $\pm$ 0.13  & -0.59  $\pm$ 0.08  & 0.90 $\pm$ 0.12
      &8.4  $\pm$ 0.7  \\
  2M17205201-2903061 & 260.216712 & -29.051722 & 155 & 4084  & 1.0  &
  -0.79 $\pm$ 0.04  & 0.49  $\pm$ 0.10  & -0.19  $\pm$ 0.06  & 0.96 $\pm$ 0.10
      &6.4  $\pm$ 1.4  \\
  2M17211817-2735530 & 260.325717 & -27.598082 & 228 & 3736  & 0.4  &
  -0.41 $\pm$ 0.03  & 0.27  $\pm$ 0.06  & -0.24  $\pm$ 0.04  & 0.60 $\pm$ 0.07
      &7.1  $\pm$ 1.1  \\
  2M17263951-2406247 & 261.664646 & -24.106882 & 126 & 4047  & 0.9  &
  -0.50 $\pm$ 0.04  &-0.00  $\pm$ 0.08  & -0.13  $\pm$ 0.05  & 0.52 $\pm$ 0.08
      &9.0  $\pm$ 2.0  \\
  2M17271907-2718040 & 261.829481 & -27.301126 & 148 & 4193  & 1.5  &
  -0.54 $\pm$ 0.04  & 0.25  $\pm$ 0.08  & -0.20  $\pm$ 0.06  & 0.55 $\pm$ 0.09
      &5.8  $\pm$ 1.3  \\
  2M17303980-2330234 & 262.665839 & -23.506523 & 240 & 3890  & 0.3  &
  -0.95 $\pm$ 0.04  & 0.19  $\pm$ 0.09  & -0.22  $\pm$ 0.06  & 0.92 $\pm$ 0.09
      &8.2  $\pm$ 0.9  \\
  2M17305251-2651528 & 262.718823 & -26.864672 & 157 & 3947  & 1.1  &
  -0.16 $\pm$ 0.03  & 0.10  $\pm$ 0.06  & -0.05  $\pm$ 0.04  & 0.73 $\pm$ 0.07
      &6.6  $\pm$ 1.4  \\
  2M17333623-2548156 & 263.400967 & -25.804361 & 147 & 4159  & 1.0  &
  -1.05 $\pm$ 0.04  & 0.56  $\pm$ 0.12  & -0.04  $\pm$ 0.08  & 0.94 $\pm$ 0.11
      &8.2  $\pm$ 1.6  \\
  2M17334208-2958347 & 263.425373 & -29.976315 & 224 & 3972  & 0.4  &
  -0.98 $\pm$ 0.04  & 0.40  $\pm$ 0.10  & -0.39  $\pm$ 0.06  & 1.07 $\pm$ 0.10
      &7.8  $\pm$ 1.1  \\
  2M17341660-2905083 & 263.569176 & -29.085642 & 332 & 3865  & 0.3  &
  -0.77 $\pm$ 0.04  & 0.43  $\pm$ 0.08  & -0.48  $\pm$ 0.05  & 1.09 $\pm$ 0.08
      &6.6  $\pm$ 0.9  \\
  2M17343610-2909472 & 263.650456 & -29.163118 & 201 & 4060  & 1.0  &
  -0.73 $\pm$ 0.04  & 0.26  $\pm$ 0.09  & -0.21  $\pm$ 0.06  & 1.01 $\pm$ 0.09
      &6.7  $\pm$ 1.5  \\
  2M17343654-1956596 & 263.652282 & -19.949903 & 119 & 4239  & 1.7  &
   0.10 $\pm$ 0.03  & 0.22  $\pm$ 0.06  & -0.43  $\pm$ 0.04  & 0.57 $\pm$ 0.06
      &5.1  $\pm$ 1.4  \\
  2M17343807-2557555 & 263.658637 & -25.965429 & 254 & 3946  & 0.0  &
  -1.26 $\pm$ 0.04  & 0.34  $\pm$ 0.12  & -0.39  $\pm$ 0.08  & 0.80 $\pm$ 0.11
      &7.7  $\pm$ 0.6  \\
  2M17350446-2932289 & 263.768624 & -29.541382 & 119 & 4247  & 1.3  &
  -0.77 $\pm$ 0.04  & 0.33  $\pm$ 0.11  & -0.23  $\pm$ 0.07  & 0.87 $\pm$ 0.10
      &8.0  $\pm$ 1.9  \\
  2M17352288-2913255 & 263.845356 & -29.22377  & 194 & 4176  & 1.0  &
  -0.97 $\pm$ 0.04  & 0.14  $\pm$ 0.11  & -0.37  $\pm$ 0.07  & 1.01 $\pm$ 0.11
      &6.2  $\pm$ 1.2  \\
  2M17353215-2759106 & 263.88397  & -27.986303 & 116 & 3921  & 0.9  &
  -0.56 $\pm$ 0.04  & 0.31  $\pm$ 0.08  &  0.13  $\pm$ 0.05  & 0.63 $\pm$ 0.08
      &9.2  $\pm$ 1.8  \\
  2M17354267-2406233 & 263.927815 & -24.106478 & 234 & 3763  & 0.5  &
  -0.71 $\pm$ 0.04  & 0.21  $\pm$ 0.07  &  0.14  $\pm$ 0.05  & 0.55 $\pm$ 0.08
      &7.6  $\pm$ 1.0  \\
  2M17382269-2748001 & 264.594549 & -27.800049 & 211 & 3877  & 0.1  &
  -1.13 $\pm$ 0.04  & 0.00  $\pm$ 0.11  & -0.36  $\pm$ 0.07  & 0.81 $\pm$ 0.11
      &9.0  $\pm$ 0.7  \\
  2M17382497-3006527 & 264.604065 & -30.114656 & 109 & 4122  & 1.3  &
  -0.85 $\pm$ 0.04  & 0.32  $\pm$ 0.11  & -0.10  $\pm$ 0.07  & 0.86 $\pm$ 0.10
      &6.0  $\pm$ 1.3  \\
  2M17390422-2943520 & 264.767613 & -29.731115 & 178 & 4047  & 0.7  &
  -1.20 $\pm$ 0.04  & 0.44  $\pm$ 0.12  & -0.35  $\pm$ 0.07  & 0.64 $\pm$ 0.11
      &7.7  $\pm$ 1.2  \\
  2M17404143-2714570 & 265.172631 & -27.249172 & 105 & 4120  & 1.3  &
  -0.77 $\pm$ 0.04  & 0.14  $\pm$ 0.10  & -0.07  $\pm$ 0.06  & 0.83 $\pm$ 0.10
      &7.9  $\pm$ 1.9  \\
  2M17415271-2715374 & 265.469643 & -27.260414 & 151 & 4187  & 1.0  &
  -1.15 $\pm$ 0.05  & 0.20  $\pm$ 0.13  & -0.35  $\pm$ 0.08  & 0.90 $\pm$ 0.12
      &7.3  $\pm$ 1.4  \\
  2M17431507-2815570 & 265.812795 & -28.26586  & 128 & 4177  & 1.5  &
  -0.86 $\pm$ 0.04  &-0.67  $\pm$ 0.11  &  0.01  $\pm$ 0.07  & 0.50 $\pm$ 0.10
      &7.5  $\pm$ 1.8  \\
  2M17442343-2627304 & 266.097638 & -26.458456 & 184 & 4030  & 0.7  &
  -0.91 $\pm$ 0.04  & 0.35  $\pm$ 0.10  & -0.22  $\pm$ 0.06  & 0.93 $\pm$ 0.10
      &8.7  $\pm$ 1.5  \\
  2M17453131-2342147 & 266.38046  & -23.704111 & 127 & 4047  & 1.1  &
  -0.50 $\pm$ 0.04  & 0.41  $\pm$ 0.08  &  0.00  $\pm$ 0.05  & 0.81 $\pm$ 0.08
      &9.6  $\pm$ 2.0  \\
  2M17464449-2531533 & 266.685384 & -25.531477 & 111 & 4132  & 1.0  &
  -0.78 $\pm$ 0.04  & 0.46  $\pm$ 0.10  &  0.08  $\pm$ 0.06  & 0.61 $\pm$ 0.10
      &9.4  $\pm$ 2.0  \\
  2M17482995-2305299 & 267.124792 & -23.091654 & 145 & 4316  & 1.3  &
  -0.92 $\pm$ 0.04  &-0.11  $\pm$ 0.12  & -0.40  $\pm$ 0.08  & 0.54 $\pm$ 0.11
      &8.0  $\pm$ 1.8  \\
  2M17494963-2318560 & 267.4568   & -23.315571 & 195 & 4069  & 0.9  &
  -0.80 $\pm$ 0.04  & 0.48  $\pm$ 0.09  & -0.34  $\pm$ 0.06  & 1.02 $\pm$ 0.09
      &6.1  $\pm$ 1.3  \\
  2M17504980-2255083 & 267.70754  & -22.91898  & 230 & 3956  & 0.8  &
  -0.57 $\pm$ 0.04  & 0.28  $\pm$ 0.07  & -0.38  $\pm$ 0.05  & 0.72 $\pm$ 0.08
      &5.5  $\pm$ 1.1  \\
  2M17514916-2859341 & 267.954859 & -28.992813 & 128 & 4152  & 0.9  &
  -1.06 $\pm$ 0.04  & 0.40  $\pm$ 0.13  & -0.30  $\pm$ 0.08  & 0.92 $\pm$ 0.12
      &7.7  $\pm$ 1.5  \\
  2M17523300-3027521 & 268.137518 & -30.464495 & 119 & 4187  & 1.0  &
  -1.38 $\pm$ 0.05  & 0.21  $\pm$ 0.17  &  0.03  $\pm$ 0.10  & 0.75 $\pm$ 0.14
      &9.3  $\pm$ 1.6  \\
  2M17524451-2830199 & 268.185495 & -28.505531 & 137 & 3879  & 0.9  &
  -0.45 $\pm$ 0.04  & 0.18  $\pm$ 0.07  &  0.10  $\pm$ 0.04  & 0.54 $\pm$ 0.07
      &7.8  $\pm$ 1.6  \\
  2M17530277-2835196 & 268.261583 & -28.588795 & 147 & 3865  & 0.2  &
  -0.81 $\pm$ 0.04  & 0.22  $\pm$ 0.09  & -0.22  $\pm$ 0.06  & 0.48 $\pm$ 0.09
      &9.6  $\pm$ 1.1  \\
  2M17534394-2826411 & 268.433095 & -28.444759 & 182 & 3811  &-0.2  &
  -1.01 $\pm$ 0.04  & 0.21  $\pm$ 0.10  & -0.22  $\pm$ 0.06  & 0.87 $\pm$ 0.10
      &10.3 $\pm$  0.7  \\
  2M17554454-2123058 & 268.93562  & -21.384953 & 159 & 4271  & 1.5  &
  -0.62 $\pm$ 0.04  & 0.24  $\pm$ 0.09  & -0.27  $\pm$ 0.06  & 0.57 $\pm$ 0.09
      &5.4  $\pm$ 1.5  \\
  2M18014817-3026237 & 270.450716 & -30.439939 & 117 & 4311  & 1.3  &
  -0.93 $\pm$ 0.04  & 0.43  $\pm$ 0.13  & -0.08  $\pm$ 0.08  & 0.88 $\pm$ 0.11
      &9.6  $\pm$ 2.1  \\
  2M18020427-1810191 & 270.517792 & -18.171999 & 241 & 3732  & 0.0  &
  -0.56 $\pm$ 0.03  & 0.11  $\pm$ 0.06  & -0.65  $\pm$ 0.04  & 0.71 $\pm$ 0.07
      &7.0  $\pm$ 0.7  \\
  2M18022530-2928338 & 270.605421 & -29.476059 & 191 & 3844  & 0.6  &
  -0.35 $\pm$ 0.03  & 0.10  $\pm$ 0.06  & -0.55  $\pm$ 0.04  & 1.08 $\pm$ 0.07
      &6.8  $\pm$ 1.3  \\
  2M18033335-2929122 & 270.888992 & -29.48674  & 95 & 4482  & 1.8  &
  -0.92 $\pm$ 0.05  & 0.82  $\pm$ 0.14  & -0.09  $\pm$ 0.09  & 0.73 $\pm$ 0.12
      &8.4  $\pm$ 2.0  \\
  2M18035944-2908195 & 270.997669 & -29.138758 & 162 & 3820  & 0.6  &
  -0.47 $\pm$ 0.03  & 0.18  $\pm$ 0.07  & -0.09  $\pm$ 0.04  & 0.57 $\pm$ 0.07
      &6.6  $\pm$ 1.1  \\
  2M18054875-3122407 & 271.453164 & -31.377975 & 389 & 3829  &-0.2  &
  -1.16 $\pm$ 0.04  & 0.09  $\pm$ 0.11  & -0.32  $\pm$ 0.07  & 0.69 $\pm$ 0.11
      &9.8  $\pm$ 0.6  \\
  2M18061336-3147053 & 271.555701 & -31.784821 & 182 & 4427  & 1.6  &
  -0.57 $\pm$ 0.04  & 0.05  $\pm$ 0.09  & -0.04  $\pm$ 0.06  & 0.51 $\pm$ 0.09
      &6.7  $\pm$ 2.1  \\
  2M18090957-1559276 & 272.289877 & -15.991026 & 135 & 3882  & 0.4  &
  -0.28 $\pm$ 0.03  & 0.04  $\pm$ 0.06  & -0.26  $\pm$ 0.04  & 0.52 $\pm$ 0.07
      &8.7  $\pm$ 1.7  \\
  2M18102953-2707208 & 272.62305  & -27.122459 & 155 & 4115  & 1.2  &
  -0.36 $\pm$ 0.04  & 0.52  $\pm$ 0.07  &  0.02  $\pm$ 0.05  & 0.58 $\pm$ 0.08
      &6.0  $\pm$ 1.4  \\
  2M18120031-1350169 & 273.001326 & -13.838031 & 131 & 4230  & 1.3  &
  -0.97 $\pm$ 0.04  & 1.16  $\pm$ 0.12  &  0.08  $\pm$ 0.08  & 0.64 $\pm$ 0.11
      &6.0  $\pm$ 1.3  \\
  2M18121957-2926310 & 273.081553 & -29.441954 & 193 & 4031  & 0.8  &
  -0.91 $\pm$ 0.04  & 0.36  $\pm$ 0.10  & -0.04  $\pm$ 0.06  & 0.62 $\pm$ 0.10
      &9.4  $\pm$ 1.7  \\
  2M18124455-2719146 & 273.185633 & -27.32074  & 211 & 3992  & 0.6  &
  -1.02 $\pm$ 0.04  & 0.09  $\pm$ 0.11  & -0.22  $\pm$ 0.07  & 0.55 $\pm$ 0.10
      &8.0  $\pm$ 1.2  \\
  2M18165340-2017051 & 274.222524 & -20.284777 & 107 & 4029  & 1.1  &
  -0.03 $\pm$ 0.03  & 0.17  $\pm$ 0.06  & -0.39  $\pm$ 0.04  & 0.66 $\pm$ 0.06
      &5.8  $\pm$ 1.6  \\
  2M18334592-2903253 & 278.441366 & -29.057034 & 154 & 4264  & 1.6  &
  -0.78 $\pm$ 0.04  & 0.65  $\pm$ 0.11  & -0.19  $\pm$ 0.07  & 0.92 $\pm$ 0.10
      &6.0  $\pm$ 1.8  \\
  2M18372953-2911046 & 279.373046 & -29.18462  & 188 & 4475  & 2.7  &
  -1.06 $\pm$ 0.04  & 0.39  $\pm$ 0.14  & -0.35  $\pm$ 0.09  & 1.00 $\pm$ 0.12
      &6.1  $\pm$ 1.3  \\
  2M18442352-3029411 & 281.098036 & -30.494764 & 287 & 4073  & 1.0  &
  -0.77 $\pm$ 0.04  & 0.73  $\pm$ 0.09  & -0.15  $\pm$ 0.06  & 0.92 $\pm$ 0.09
      &5.4  $\pm$ 1.1  \\
  2M18550318-3043368 & 283.763269 & -30.726915 & 71 & 4444  & 2.1  &
  -0.93 $\pm$ 0.04  & 1.06  $\pm$ 0.15  & -0.13  $\pm$ 0.09  & 0.97 $\pm$ 0.13
      &10.1 $\pm$  2.3  \\
\hline
\end{tabular}
\end{adjustbox}
\label{expectednumbers}
\end{table*}

\label{lastpage}

\end{document}